\documentclass[11pt,twoside,letterpaper]{article} %% The same for {book}
\usepackage{times,fancyhdr,graphicx}
\usepackage[utf8]{inputenc}
\usepackage[T1]{fontenc}
\usepackage{url}
\usepackage{makeidx}
\makeindex
\def\2F1{~_2F_1}
\def\sun{\hbox{$\odot$}}
\def\jcap{Journal of Cosmology and Astroparticle Physic  } %Journal of Cosmology and Astroparticle Physics
\def\mnras{MNRAS\,  }%% Monthly Notices of the RAS

\def\prd{Phys. Rev. D   }% % Physical Review D
  

\makeatletter
\def\cleardoublepage{\clearpage\if@twoside \ifodd\c@page\else%
    \hbox{}%
    \thispagestyle{empty}%
    \newpage%
    \if@twocolumn\hbox{}\newpage\fi\fi\fi}
\makeatother

\renewcommand{\thesection}{\arabic{section}.}
\renewcommand{\thesubsection}{\thesection\arabic{subsection}.}
\renewcommand{\thesubsubsection}{\thesubsection\arabic{subsubsection}.}

\def\figurename{Figure}
\makeatletter
\renewcommand{\fnum@figure}[1]{\figurename~\thefigure.}
\makeatother

\def\tablename{Table}
\makeatletter
\renewcommand{\fnum@table}[1]{\bf\tablename~\thetable.}
\makeatother

\begin{document}
\title{
{\begin{flushleft} \vskip 0.45in
{\normalsize\bfseries\textit{Chapter~}}
\end{flushleft}
\vskip 0.45in \bfseries\scshape
Conservation of the Flux of  Energy   in  Extra-Galactic  Jets
}}
\author{\bfseries\itshape
 Lorenzo Zaninetti \thanks{Author's Email: zaninetti@ph.unito.it}\\
Physics Department, Turin,Italy
}

\date{}
\maketitle \thispagestyle{empty} \setcounter{page}{1}
%% ------- [First Page Running Head] - place it immediately after title! ------
%\thispagestyle{fancy} \fancyhead{}
%\fancyhead[L]{In: Book Title \\
%Editor: Editor Name, pp. {\thepage-\pageref{lastpage-01}}} % needs \label{lastpage-01} on the last page.
%\fancyhead[R]{ISBN 0000000000  \\
%\copyright~2007 Nova Science Publishers, Inc.} \fancyfoot{}
%\renewcommand{\headrulewidth}{0pt}
%%------------------------------------------------------------------------------

%\vspace{2in}
%
%\noindent{\it Keywords}:
%Galaxies: jets
%Relativity

%\section*{Abstract}
\begin{abstract}
The conservation of the energy flux  in turbulent jets
that propagate in the intergalactic medium (IGM)
allows us to deduce the  law of motion
in the classical and relativistic cases.
Four types of IGM are considered: constant density,
hyperbolic decrease of density, inverse power law decrease of density
and  a Lane--Emden ($n=5$) profile.
The conservation of the relativistic
flux for the   energy
allows us to derive, to the first order,  an analytical
expression for the velocity. It also allows us to
numerically determine the trajectory for the four types of medium.
In the case of a Lane--Emden ($n=5$) profile,
the back-reaction due to the radiative losses for the trajectory
is evaluated  both in the classical and the relativistic case.
Astrophysical applications are made to the centerline
intensity of the synchrotron emission
and to the evolution of the magnetic field
in the case of the radio-galaxy 3C31.
\end{abstract}

\vspace{.08in}

% ------------ [Running Heads - for odd and even pages] - please insert it only on page 2!
\pagestyle{fancy}
\fancyhead{}
\fancyhead[EC]{\it  Lorenzo Zaninetti }
\fancyhead[EL,OR]{\thepage}
\fancyhead[OC]{\it Conservation of the Flux of Energy ... }
\fancyfoot{}
\renewcommand\headrulewidth{0.0pt}
%------------------------------------------------------------------------------

\section{Introduction}
The analysis  of turbulent jets in the laboratory
offers the possibility of applying the theory  of turbulence
to some well-defined  experiments, see
\cite{Reynolds1883,Reynolds1894}.
Reynolds experiments can be seen
in \cite{vanDyke1982}.
Analytical results for the theory  of turbulent jets
can be found  in \cite{goldstein,landau,pope2000,foot}.
Recently, the analogy between laboratory
jets and extra-galactic radio-jets has been pointed out,
see
\cite{Lebedev2011,Suzuki-Vidal2012}.
We briefly recall that the theory of
`round turbulent jets' can be defined in terms
of the velocity at the nozzle, the diameter of the nozzle,
and the viscosity, see Section 5 in \cite{pope2000}.
However, in this example, the gradients in pressure are not considered.
The application of the theory of turbulence to extra-galactic
radio-jets raises  many questions
because we do not observe the turbulent phenomena,
but the radio features that have properties
similar to the laboratory's turbulent jets, i.e., similar opening angles.
We now  pose
the following questions:
\begin{itemize}
\item Is it possible to apply the conservation of the flux
     of energy to derive the equation of
      motion for radio-jets in the cases of constant
     and variable density of
     the surrounding medium?
\item Can we extend the conservation of the flux of energy
    to the relativistic regime?
\item Can we model the behaviour of the magnetic field
      and the intensity of synchrotron emission as functions
      of the distance from the parent nucleus?
\item Can we model the back reaction on the equation of motion
      for turbulent jets due to radiative losses?
\end{itemize}
To answer these questions,
in  Sections
\ref{secclassic} and
\ref{secrelativistic},  we derive
the  differential equations
that model the classical and relativistic
conservation of the energy flux
for a  turbulent jet in  the presence of different
types of medium.
%modificare
Sections \ref{secclassiclosses}   and \ref{secrelativisticlosses} present
the classical and the relativistic parametrization
of the radiative losses
for the Lane--Emden ($n=5$) profile.
Section \ref{secapplication}
introduces
two models for the synchrotron emission
along the jet.

\section{Energy Conservation}
\label{secclassic}

The conservation of the energy    flux in a turbulent jet
requires a perpendicular section to the motion along the
Cartesian $x$-axis, $A$
\begin {equation}
A(r)=\pi~r^2
\end{equation}
where $r$ is the radius of the jet.
Section  $A$ at  position $x_0$  is
\begin {equation}
A(x_0)=\pi ( x_0   \tan ( \frac{\alpha}{2}))^2
\end{equation}
where   $\alpha$  is the opening angle and
$x_0$ is the initial position on the $x$-axis.
At position $x$, we have
\begin {equation}
A(x)=\pi ( x   \tan ( \frac{\alpha}{2}))^2
\quad .
\end{equation}
The conservation  of energy flux states that
\begin{equation}
\frac{1}{2} \rho(x_0)  v_0^3   A(x_0)  =
\frac{1}{2} \rho(x  )   v(x)^3 A(x)
\label{conservazioneenergy}
\end {equation}
\index{conservation~energy~flux!classic}

\noindent\noindent where $v(x)$ is the velocity at  position $x$ and
$v_0(x_0)$   is the velocity at  position $x_0$,
see Formula A28 in \cite{deyoung}.

The selected physical units are
pc for length  and  yr for time;
with these units, the initial velocity $v_{{0}}$
is  expressed in pc yr$^{-1}$,
 1 yr = 365.25 days.
When the initial velocity is expressed in
km\,s$^{-1}$, the multiplicative factor $1.02\times10^{-6}$
should be applied in order to have the velocity expressed in
pc yr$^{-1}$.
More details can be found in \cite{Zaninetti2016e,Zaninetti2018b}

\subsection{Constant  Density}

\label{classicalconstant}
\index{profile~of~density!constant}
In the case of constant density of the intergalactic medium (IGM)
along the $x$-direction,
the  law of conservation of the
energy flux, as given by Eq. (\ref{conservazioneenergy}),
can be written as a  differential equation
\begin{equation}
\left( {\frac {\rm d}{{\rm d}t}}x \left( t \right)  \right) ^{3}
 \left( x \left( t \right)  \right) ^{2}-{v_{{0}}}^{3}{x_{{0}}}^{2}
=0
\quad .
\label{diffequationclassic}
\end{equation}
The analytical  solution of the previous differential
equation can be found by imposing $x=x_0$ at t=0,
\begin{equation}
x(t) =
\frac{1}{3}\,{3}^{2/5}\sqrt [5]{{x_{{0}}}^{2} \left( 5\,tv_{{0}}+3\,x_{{0}}
 \right) ^{3}}
\quad .
\label{energyxtconstant}
\end{equation}
The asymptotic approximation
is
\begin{equation}
x(t) \sim
\frac{1}{3}\,{3}^{2/5}{5}^{3/5}\sqrt [5]{{v_{{0}}}^{3}{x_{{0}}}^{2}}{t}^{3/5}
\quad .
\end{equation}
The velocity is
\begin{equation}
v(t) =
{\frac {{3}^{2/5}{x_{{0}}}^{2} \left( 5\,tv_{{0}}+3\,x_{{0}} \right) ^
{2}v_{{0}}}{ \left( {x_{{0}}}^{2} \left( 5\,tv_{{0}}+3\,x_{{0}}
 \right) ^{3} \right) ^{4/5}}}
\label{energyvtconstant}
\end{equation}
and its asymptotic approximation
\begin{equation}
v(t) \sim
\frac{1}{5}\,{\frac {{3}^{2/5}\sqrt [5]{125}{x_{{0}}}^{2}{v_{{0}}}^{3} \left(
{t}^{-1} \right) ^{2/5}}{ \left( {v_{{0}}}^{3}{x_{{0}}}^{2} \right) ^{
4/5}}}
\quad  .
\end{equation}
The velocity as a function of the distance is
\begin{equation}
v(x) = {\frac {{x_{{0}}}^{2/3}v_{{0}}}{{x}^{2/3}}}
\quad .
\label{energyvx}
\end{equation}

A first comparison can be made  with the laboratory
data on turbulent jets of  \cite{Mistry2014}
where the  velocity of the turbulent jet
at the nozzle diameter, $D_j$=1,
is $v_0 =2.53$\ m s$^{-1}$
and at  $D_j$=50 the  centerline  velocity
is $v =0.314$\ m s$^{-1}$.
The formula (\ref{energyvx})  with $x_0=1$ and $x=50$
gives  an averaged velocity of $v=0.186$\ m s$^{-1}$
which multiplied by 2 gives $v =0.372$\ m s$^{-1}$.
This multiplication by 2 has been done because
the turbulent jet  develops a profile of velocity in the direction
perpendicular  to the jet's main axis and, therefore,
the centerline velocity is  approximately double that of the
averaged velocity.
The transit time, $t_{tr}$, necessary to travel a distance
of $x_{max}$ can be derived from Eq. (\ref{energyxtconstant})
\begin{equation}
t_{tr} =
\frac
{
3\,\sqrt [3]{{x_{{\max}}}^{2}x_{{0}}}x_{{\max}}-3\,{x_{{0}}}^{2}
}
{
5\,x_{{0}}v_{{0}}
}
\quad .
\label{energytransit}
\end{equation}

An astrophysical  test can be  performed on a typical distance
of 15 kpc relative  to the jets in 3C\,31,
see Figure 2 in \cite{laing2002}.
On inserting  $x=15000\,$pc$=15$\ kpc, $x_0=100$\ pc,
and $v_0=10000$\ km s$^{-1}$ we obtain a transit time
of  $t_{tr}=2.488\,10^7$\ yr.

The rate of mass flow at the point $x$, $\dot {m}(x)$, is
\begin{equation}
\dot {m}(x) =
\rho   v(x)  \pi ( x   \tan ( \frac{\alpha}{2}))^2
\end{equation}
and the astrophysical version is
\begin{equation}
\dot {m}(x) =
0.0237n {x}^{4/3} \left( \tan
 \left( \alpha/2 \right)  \right) ^{2}{x_{{0}}}^{2/3}\beta_{{0}}\,\frac{\it
M_{\sun}}{yr}
\end{equation}
where $x$ and $x_0$ are expressed in pc,
$n $ is the number density of protons   expressed  in
particles~cm$^{-3}$,
$M_{\sun}$ is the solar mass
and $\beta_0=\frac{v_0}{c}$.
The previous formula indicates  that
the rate of transfer of particles
is not constant along the
jet but increases  $\propto x^{4/3}$.

\subsection{A Hyperbolic Profile of the Density}

Now the density  is  assumed to decrease as
\begin{equation}
\rho = \rho_0  (\frac{x_0}{x})
\label{profhyperbolic}
\end{equation}
\index{profile~of~density!hyperbolic}

\noindent where  $\rho_0=0$ is the density at  $x=x_0$.
The differential equation that models the
energy  flux is
\begin{equation}
x_{{0}}x \left( t \right)  \left( {\frac {\rm d}{{\rm d}t}}x \left( t
 \right)  \right) ^{3}-{v_{{0}}}^{3}{x_{{0}}}^{2}
= 0
\end{equation}
and its analytical solution
is
\begin{equation}
x(t) =
\frac{1}{3}\,\sqrt [4]{3}\sqrt [4]{x_{{0}} \left( 4\,tv_{{0}}+3\,x_{{0}}
 \right) ^{3}}
\quad .
\label{xthyperbolic}
\end{equation}
The asymptotic  approximation
is
\begin{equation}
x(t)  \sim
\frac{2}{3} \,\sqrt [4]{3}\sqrt {2}\sqrt [4]{{v_{{0}}}^{3}x_{{0}}}{t}^{3/4}
\quad .
\label{xthyperbolicasympt}
\end{equation}
The analytical solution for the velocity
is
\begin{equation}
v(t) =
{\frac {\sqrt [4]{3}x_{{0}} \left( 4\,tv_{{0}}+3\,x_{{0}} \right) ^{2}
v_{{0}}}{ \left( x_{{0}} \left( 4\,tv_{{0}}+3\,x_{{0}} \right) ^{3}
 \right) ^{3/4}}}
\label{vthyperbolic}
\end{equation}
and its asymptotic  approximation is
\begin{equation}
v(t) \sim
\frac{1}{4}\,{\frac {\sqrt [4]{3}\sqrt [4]{64}x_{{0}}{v_{{0}}}^{3}\sqrt [4]{{t
}^{-1}}}{ \left( {v_{{0}}}^{3}x_{{0}} \right) ^{3/4}}}
\quad  .
\end{equation}

The transit time can be  derived from
Eq. (\ref{xthyperbolic})
\begin{equation}
t_{tr} =
\frac
{
3\,\sqrt [3]{x_{{\max}}{x_{{0}}}^{2}}x_{{\max}}-3\,{x_{{0}}}^{2}
}
{
4\,x_{{0}}v_{{0}}
}
\end{equation}
and with $x=15000$\ pc$=15$\ kpc, $x_0=100$\ pc,
and $v_0=10000$\ km s$^{-1}$  as in Section
\ref{classicalconstant},
we have  $t_{tr} =5.848\,10^6 $\ yr.

\subsection{An Inverse Power Law  Profile of the Density}

\begin{figure*}
\begin{center}
\includegraphics[width=3cm,angle=-90]{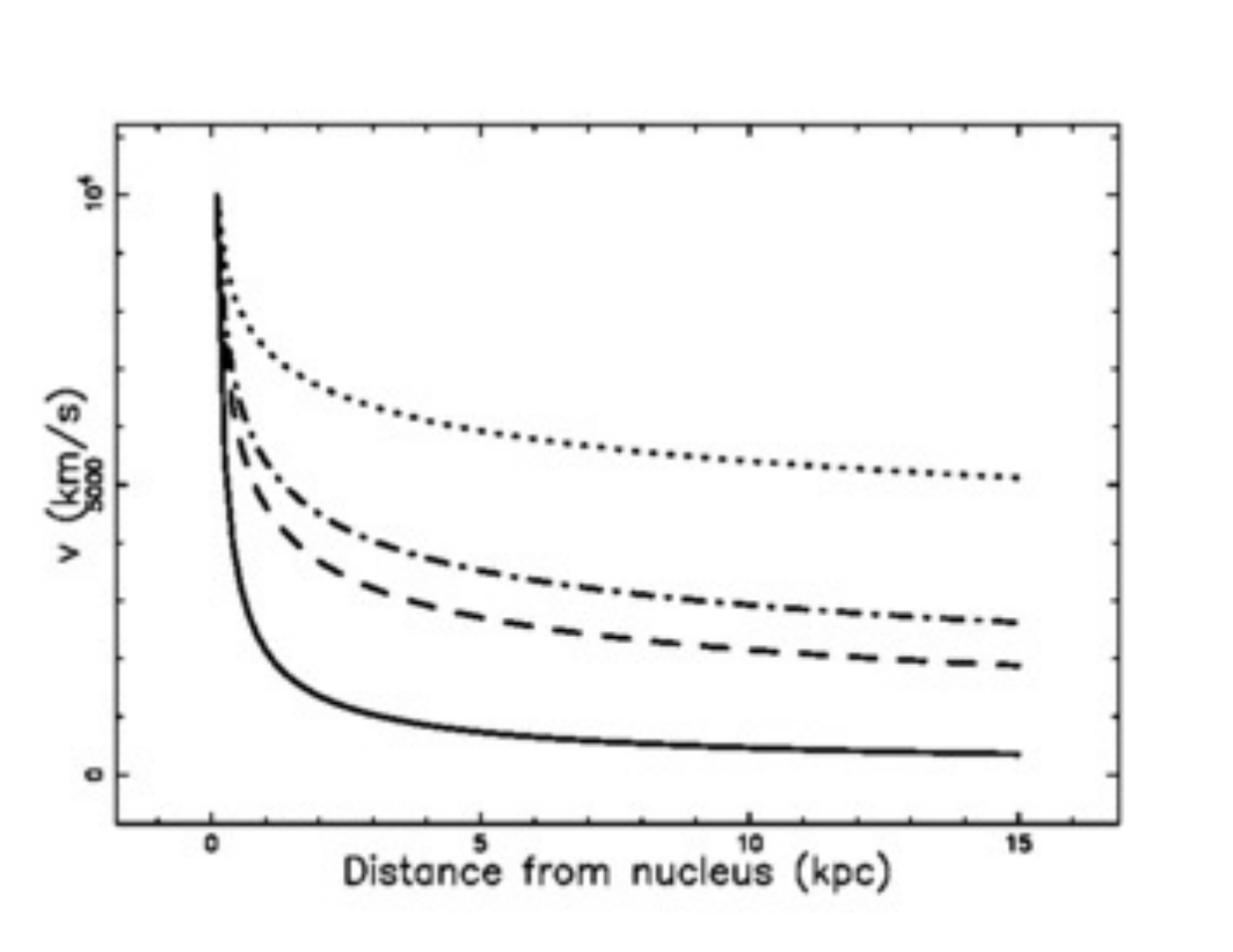}
\end {center}
\caption
{
Classical velocity   as a function
of  the distance from the nucleus  when
$x_0$ =100~pc and $v_0=10000\,$km s$^{-1}$:
$\delta =0$   (full line),
$\delta =1$   (dashes),
$\delta =1.2$ (dot-dash-dot-dash)
and
$\delta =1.6$ (dotted).
}
\label{energyveldelta}
    \end{figure*}

Here, the density  is  assumed to decrease as
\begin{equation}
\rho = \rho_0  (\frac{x_0}{x})^{\delta}
\label{profpower}
\end{equation}
where  $\rho_0$ is the density at  $x=x_0$.
The differential equation which  models the
energy  flux is
\begin{equation}
1/2\, \left( {\frac {x_{{0}}}{x}} \right) ^{\delta} \left( {\frac 
{\rm d}{{\rm d}t}}x \left( t \right)  \right) ^{3}{x}^{2}-1/2\,{v_{{0}
}}^{3}{x_{{0}}}^{2}
=0
\quad  .
\end{equation}
There is no analytical solution, and we simply express
the velocity as a function of the position, $x$,
\begin{equation}
v(x) =
\frac
{
\sqrt [3]{{x_{{0}}}^{2} \left(  \left( {\frac {x_{{0}}}{x}} \right) ^{
\delta} \right) ^{2}x}v_{{0}}
}
{
\left( {\frac {x_{{0}}}{x}} \right) ^{\delta}x
}
\label{velocitypower}
\end{equation}
see  Figure \ref{energyveldelta}
%citiamofigura_energyveldelta
% figure   energyveldelta
% end energyveldelta

The rate of mass flow at the point $x$ is
\begin{equation}
\dot {m}(x) =
\rho_{{0}}\pi\,x \left( \tan \left( \alpha/2 \right)  \right) ^{2}
\sqrt [3]{{x_{{0}}}^{2} \left(  \left( {\frac {x_{{0}}}{x}} \right) ^{
\delta} \right) ^{2}x}v_{{0}}
\label{mxpower}
\end{equation}
and the astrophysical version is
\begin{equation}
\dot {m}(x) =
0.02375\,nx \left( \tan \left( \alpha/2 \right)  \right) ^{2}
\sqrt [3]{{x_{{0}}}^{2} \left(  \left(  1.0\,{\frac {x_{{0}}}{x}}
 \right) ^{\delta} \right) ^{2}x}\beta_{{0}}{\it Msun}
\end{equation}
where
$n_0 $ is the number density of protons   expressed  in
particles~cm$^{-3}$ at $x_0$.

\subsection{The Lane--Emden Profile}

\label{seclaneclassic}
The self-gravitating sphere of a polytropic
gas is governed
by the Lane--Emden differential equation
of the second order
\begin{equation}
{\frac {d^{2}}{d{x}^{2}}}Y ( x ) +2\,{\frac {{\frac {d}{dx}
}Y ( x ) }{x}}+ ( Y ( x )  ) ^{n}=0
\quad,
\nonumber
\end{equation}
where $n$  is an integer, see
\cite{Lane1870, Emden1907, Chandrasekhar_1967, Binney2011, Zwillinger1989}.
The solution  $Y ( x )_n$
has  the density profile
\begin{equation}
\rho = \rho_c Y ( x )_n^n
\quad,
\nonumber
\end{equation}
where $\rho_c$ is the density at $x=0$.
The pressure $P$ and temperature $T$ scale as
\begin{equation}
P = K \rho^{1 +\frac{1}{n}}
\quad,
\label{pressure}
\end{equation}
\begin{equation}
 T = K^{\prime} Y(x)
\label{temperature}
\quad,
\end{equation}
where  K and K$^{\prime}$   are two  constants.
For more details, see \cite{Hansen1994}.

Analytical solutions exist for $n=0$, 1, and 5.
The analytical  solution for $n$=5 is
\begin{equation}
Y(x) ={\frac {1}{{(1+ \frac{{x}^{2}}{3})^{1/2}}} }
\quad,
\nonumber
\end{equation}
and the density for $n$=5 is
\begin{equation}
\rho(x) =\rho_c {\frac {1}{{(1+ \frac{{x}^{2}}{3})^{5/2}}} }
\label{densitan5}
\quad.
\end{equation}

The   variable  $x$   is   non-dimensional
and  we now  introduce the
new variable $x=r/b$
\begin{equation}
\rho(r;b) =\rho_c {\frac {1}{{(1+ \frac{{r}^{2}}{3b^2})^{5/2}}} }
\label{densita5b}
\quad .
\end{equation}
\index{profile~of~density!Lane--Emden}
Then,
the   conservation of the flux of energy
is
\begin{eqnarray}
\frac{1}{2}\,{\rho_{{0}}{v(x)}^{3}\pi\,{x}^{2} \left( \tan \left( \frac{\alpha}{2}
 \right)  \right) ^{2} \left( 1+\frac{1}{3}\,{\frac {{x}^{2}}{{b}^{2}}}
 \right) ^{-5/2}}
\nonumber \\
=\frac{1}{2}\,{\rho_{{0}}{v_{{0}}(x_0)}^{3}\pi\,{x_{{0}}}^{2}
 \left( \tan \left( \frac{\alpha}{2} \right)  \right) ^{2} \left( 1+\frac{1}{3}\,{
\frac {{x_{{0}}}^{2}}{{b}^{2}}} \right) ^{-5/2}}
\quad ,
\end{eqnarray}
where $v(x)$ is the velocity at  position $x$,
$v_0(x_0)$   is the velocity at  position $x_0$
and
$\alpha$  is the opening angle of the jet.
This equation is a cubic equation,
which
has  one real root plus
two non-real complex conjugate roots.
Here, and in the following, we only take the real root  into account.
The  real analytical  solution for the velocity without losses is
\begin{equation}
v(x;b,x_0,v_0) =
\frac
{
v_{{0}} \left( 3\,{b}^{2}+{x}^{2} \right) ^{{\frac{5}{6}}}{x_{{0}}}^{{
\frac{2}{3}}}
}
{
 \left( 3\,{b}^{2}+{x_{{0}}}^{2} \right) ^{{\frac{5}{6}}}{x}^{{\frac{2
}{3}}}
}
\quad .
\label{vfirst}
\end{equation}
The asymptotic expansion of above velocity, $v_a$,
with respect to the variable $x$, which means $x\rightarrow \infty$,
is
\begin{equation}
v_a(x;b,x_0,v_0)
=
\frac
{
v_{{0}}{x_{{0}}}^{{\frac{2}{3}}} \left( 5\,{b}^{2}+2\,{x}^{2} \right)
}
{
2\, \left( 3\,{b}^{2}+{x_{{0}}}^{2} \right) ^{5/6}x
}
\label{vfirstasymptotic}
\quad .
\end{equation}
The trajectory can be found by the indefinite
integral of
the inverse of the velocity as  given by equation (\ref{vfirst}):
\begin{equation}
F(x)=\int   \frac{1}{v(x;b,x_0,v_0)} dx
=
\frac
{
\sqrt [6]{3} \left( 3\,{b}^{2}+{x_{{0}}}^{2} \right) ^{{\frac{5}{6}}}{
x}^{{\frac{5}{3}}}
{\mbox{$_2$F$_1$}({\frac{5}{6}},{\frac{5}{6}};\,{\frac{11}{6}};\,-{\frac
{{x}^{2}}{3\,{b}^{2}}})}
}
{
5\,v_{{0}} \left( {b}^{2} \right) ^{5/6}{x_{{0}}}^{2/3}
}
\quad ,
\end{equation}
where ${\2F1(a,b;\,c;\,v)}$ is a regularized hypergeometric function,
see \cite{Abramowitz1965,Seggern1992,Thompson1997,NIST2010}.
The trajectory expressed in terms of  $t$ as a function  of $x$
is
\begin{equation}
F(x) - F(x_0) = t
\quad  .
\label{xt}
\end{equation}
This equation cannot be inverted in the usual form,
which is  $x$ as a function of $t$.
The asymptotic trajectory can be found by the indefinite
integral of
the inverse of the asymptotic  velocity as  given by
equation (\ref{vfirstasymptotic})
\begin{equation}
F_a(x)=\int   \frac{1}{v_a(x;b,x_0,v_0)} dx
=
\frac
{
 \left( 3\,{b}^{2}+{x_{{0}}}^{2} \right) ^{{\frac{5}{6}}}\ln  \left( 5
\,{b}^{2}+2\,{x}^{2} \right)
}
{
2\,v_{{0}}{x_{{0}}}^{2/3}
}
\quad .
\end{equation}
The equation of the asymptotic trajectory is
\begin{equation}
F_a(x) - F_a(x_0) = t
\quad  ,
\end{equation}
and the solution for $x$ of the above equation,
the asymptotic trajectory, is
\begin{equation}
x(t;b,x_0,v_0)=
\frac{1}{2}
\sqrt {-10\,{b}^{2}+2\,{{\rm e}^{{\frac { \left( 3\,{b}^{2}+{x_{{0}}}^
{2} \right) ^{5/6}\ln  \left( 5\,{b}^{2}+2\,{x_{{0}}}^{2} \right) +2\,
tv_{{0}}{x_{{0}}}^{2/3}}{ \left( 3\,{b}^{2}+{x_{{0}}}^{2} \right) ^{5/
6}}}}}}
\quad .
\label{xtasymptotic}
\end{equation}
\

\begin{figure*}
\begin{center}
\includegraphics[width=7cm]{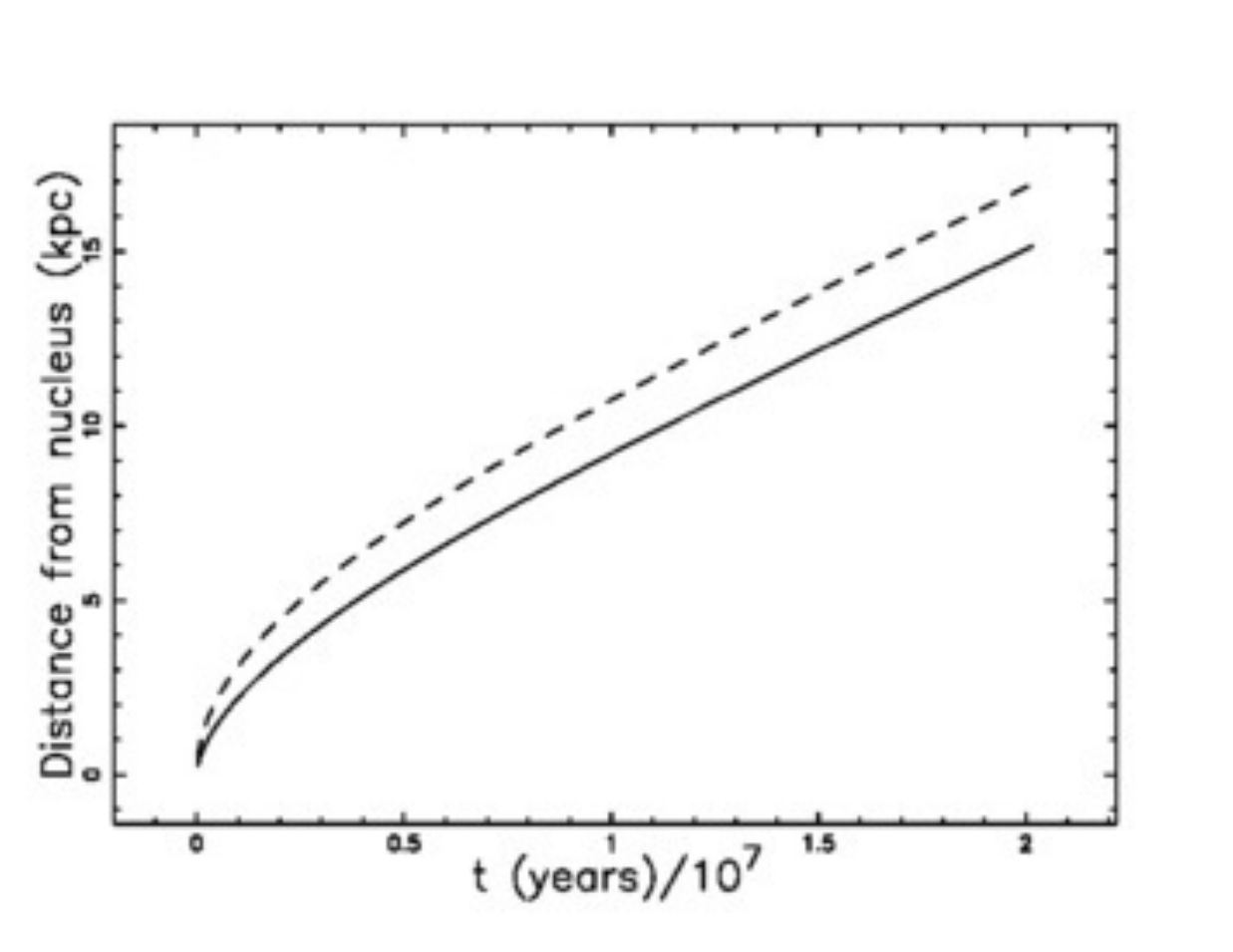}
\end {center}
\caption
{
The numerical solution  for a  Lane--Emden profile as given
by equation (\ref{xt}) (full line)
and asymptotic  solution
 as given
by equation (\ref{xtasymptotic}) (dashed line),
with parameters as in Table \ref{jetparameters}.
}
\label{traj_asymp}
    \end{figure*}

\noindent Figure  \ref{traj_asymp} shows a typical example
of the above asymptotic  expansion.

%begin table jetparameters
\begin{table}[ht!]
\caption
{
Parameters for a classical extra-galactic jet
}
\label{jetparameters}
\begin{center}
\begin{tabular}{|c|c|}
\hline
parameter    &  value    \\
\hline
$x_0$ (pc)   & 100       \\
$v_0$ ($\frac{km}{s}$)   & 10000       \\
$b$   (pc)   & 10000\\
\hline
\end{tabular}
\end{center}
\end{table}
%end  table jetparameters

%figure traj_asymp
% end figure traj_asymp

\subsection{Solution to Second Order for the Lane--Emden Profile}

\label{secclassiclosses}

Let us suppose that the radiative losses
in the case of a  Lane--Emden profile
are  proportional to the   flux of energy
\begin{equation}
- \epsilon
\frac
{
\rho_{{0}}{v}^{3}\pi\,{x}^{2} \left( \tan \left( {\frac {\alpha}{2}}
 \right)  \right) ^{2}
}
{
2\, \left( 1+\frac{1}{3}\,{\frac {{x}^{2}}{{b}^{2}}} \right) ^{5/2}
}
\quad .
\end{equation}

By inserting in the above equation  the  velocity to first order
as  given by equation~(\ref{vfirst}),
the radiative losses, $Q(x;x_0,v_0,b,\epsilon)$, are
\begin{equation}
Q(x;x_0,v_0,b,\epsilon)= - \epsilon
\frac
{
\rho_{{0}}{v}^{3}\pi\,{x}^{2} \left( \tan \left( {\frac {\alpha}{2}}
 \right)  \right) ^{2}
}
{
2\, \left( 1+\frac{1}{3}\,{\frac {{x}^{2}}{{b}^{2}}} \right) ^{5/2}
}
\quad ,
\label{lossesclassical}
\end{equation}

\noindent where $\epsilon$ is a constant which  fixes   the conversion
of  the   flux of energy  to other kinds of energies; in this
case, the radiative losses.
The sum of the radiative  losses between $x_0$ and $x$
is given by the following integral, $L$,

{\footnotesize
\begin{equation}
L(x;x_0,v_0,b,\epsilon)=\int_{x_0}^x  Q(x;x_0,v_0,b,\epsilon) dx
=\frac
{
-9\,\epsilon\,\rho_{{0}}\sqrt {3}{b}^{5}{v_{{0}}}^{3}{x_{{0}}}^{2}\pi
\, \left( \tan \left( \alpha/2 \right)  \right) ^{2} \left( x-x_{{0}}
 \right)
}
{
2\, \left( 3\,{b}^{2}+{x_{{0}}}^{2} \right) ^{5/2}
}.
\label{classiclosses}
\end{equation}
}

\noindent  The  conservation of the   flux of energy  in the presence
of  the back-reaction due to the radiative losses
for the    Lane--Emden profile
is

{\small
\begin{eqnarray}
&\frac
{
9\,\sqrt {3}\rho_{{0}} \left( {b}^{5}{v_{{0}}}^{3}{x_{{0}}}^{2}
\epsilon\, \left( {\frac {3\,{b}^{2}+{x}^{2}}{{b}^{2}}} \right) ^{5/2}
x-{b}^{5}{v_{{0}}}^{3}{x_{{0}}}^{3}\epsilon\, \left( {\frac {3\,{b}^{2
}+{x}^{2}}{{b}^{2}}} \right) ^{5/2}+{v}^{3}{x}^{2} \left( 3\,{b}^{2}+{
x_{{0}}}^{2} \right) ^{5/2} \right)
}
{
2\, \left( {\frac {3\,{b}^{2}+{x}^{2}}{{b}^{2}}} \right) ^{5/2}
 \left( 3\,{b}^{2}+{x_{{0}}}^{2} \right) ^{5/2}
}
\nonumber \\
&=
{
9\,\rho_{{0}}\sqrt {3}{v_{{0}}}^{3}{x_{{0}}}^{2}
}
{
2\, \left( {\frac {3\,{b}^{2}+{x_{{0}}}^{2}}{{b}^{2}}} \right) ^{5/2}
}.
\label{consfluxback}
\end{eqnarray}
}

\begin{figure*}
\begin{center}
\includegraphics[width=8cm]{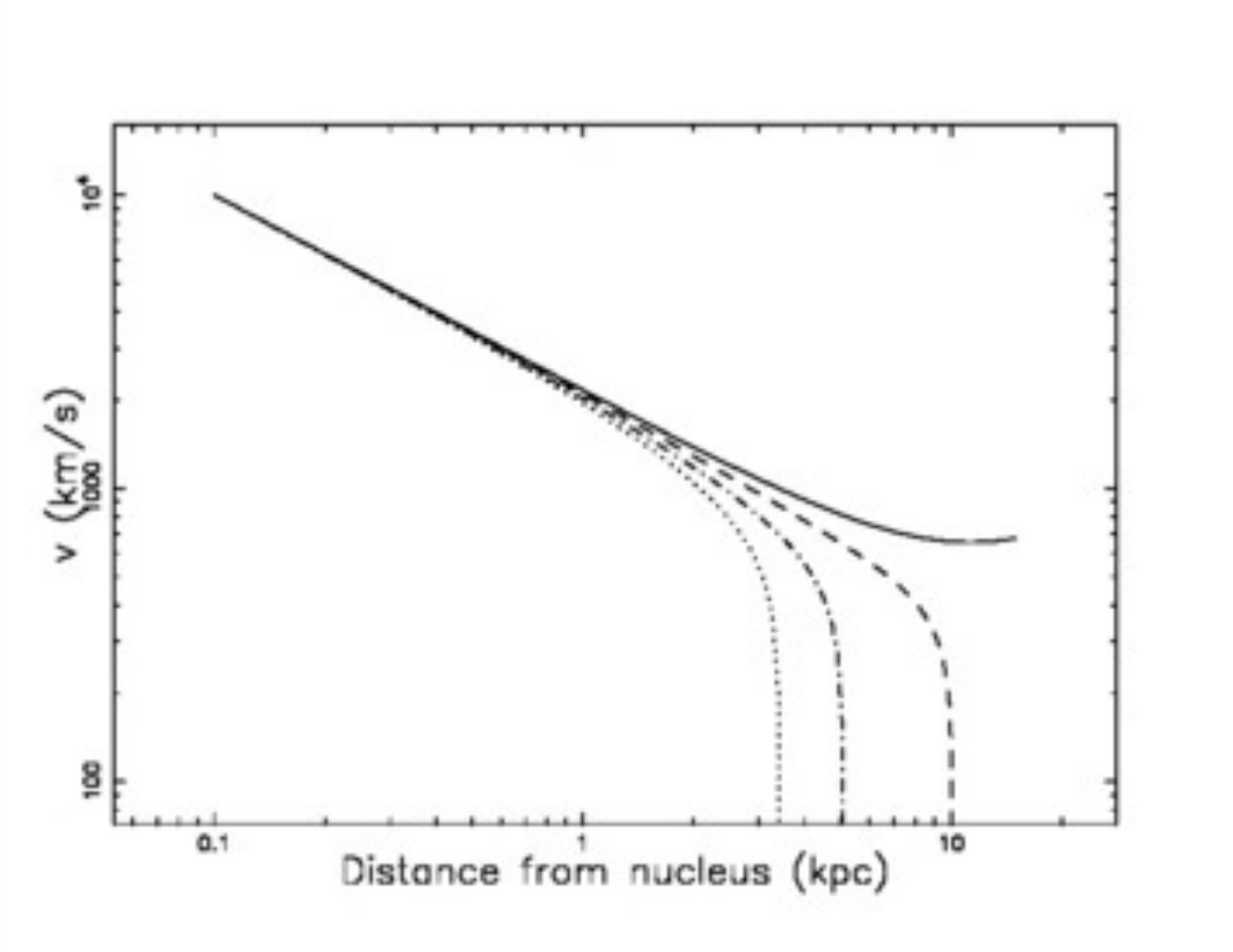}
\end {center}
\caption
{
Velocity corrected for radiative losses
for the    Lane--Emden profile, i.e., velocity to second order,
equation (\ref{vcorrected}), as a function of the distance,
with parameters as in Table \ref{jetparameters}:
$\epsilon=0 $ full   line,
$\epsilon=1.0\,10^{-4}$ dashed line,
$\epsilon=2.0\,10^{-4}$ dot-dash-dot-dash line
and
$\epsilon=3.0\,10^{-4}$ dotted line.
}
\label{vel_emden_back}
    \end{figure*}

\noindent \noindent The  analytical solution for the velocity to
second order, $v_c(x;b,x_0,v_0)$,
for the    Lane--Emden profile
is
\begin{equation}
v_c(x;b,x_0,v_0)=
\frac
{
v_{{0}}\sqrt [3]{1+\epsilon\, \left( -x+x_{{0}} \right) } \left( 3\,{b
}^{2}+{x}^{2} \right) ^{{\frac{5}{6}}}{x_{{0}}}^{{\frac{2}{3}}}
}
{
\left( 3\,{b}^{2}+{x_{{0}}}^{2} \right) ^{{\frac{5}{6}}}{x}^{{\frac{2
}{3}}}
}
\label{vcorrected}
\quad ,
\end{equation}
and  Figure \ref{vel_emden_back}  shows
an example.

%figure vel_emden_back
% end figure vel_emden_back

\begin{figure*}
\begin{center}
\includegraphics[width=8cm]{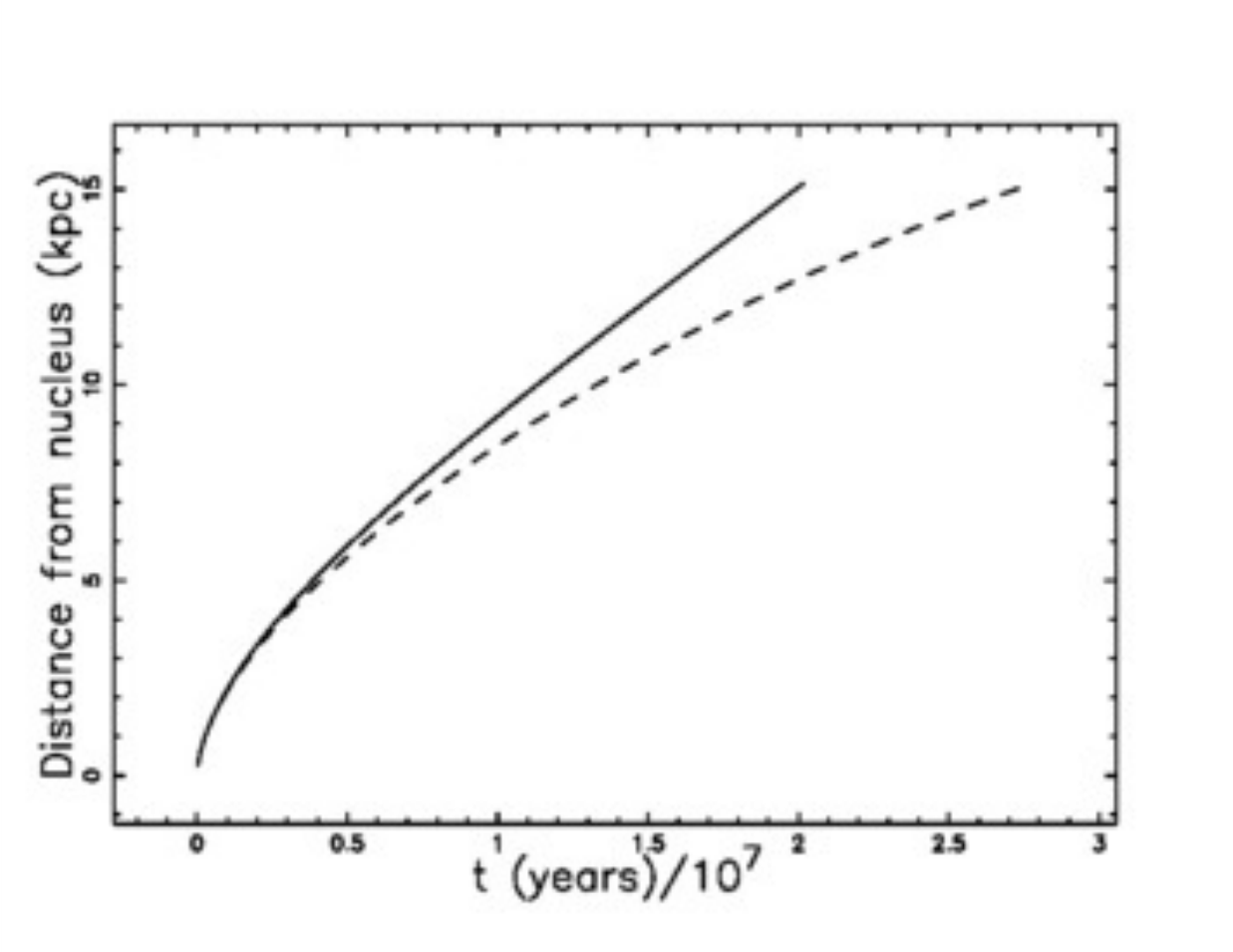}
\end {center}
\caption
{
Numerical trajectory
for a    Lane--Emden profile
corrected for radiative losses
as a function of time,
with parameters as in Table \ref{jetparameters}:
$\epsilon=0 $ full   line and
$\epsilon=8.0\,10^{-5}$ dashed line.
}
\label{traj_back}
    \end{figure*}

There are no analytical  results for the
trajectory corrected for radiative losses
for the    Lane--Emden profile,
a numerical example is shown
in Figure \ref{traj_back}.

%figure traj_back
% end figure traj_back

The  inclusion  of back-reaction  allows the evaluation of the
jet's length,  which can be derived from the minimum
in the corrected velocity to second order as a function of $x$,
\begin{equation}
\frac{\partial v_c(x;b,x_0,v_0)}{\partial x} =0
\quad ,
\end {equation}
which is
\begin{eqnarray}
-{\frac {v_{{0}}\epsilon}{3} \left( 3\,{b}^{2}+{x}^{2} \right) ^{{
\frac{5}{6}}}{x_{{0}}}^{{\frac{2}{3}}} \left( 1+\epsilon\, \left( -x+x
_{{0}} \right)  \right) ^{-{\frac{2}{3}}} \left( 3\,{b}^{2}+{x_{{0}}}^
{2} \right) ^{-{\frac{5}{6}}}{x}^{-{\frac{2}{3}}}}
\nonumber \\
+{\frac {5\,v_{{0}}
}{3}\sqrt [3]{1+\epsilon\, \left( -x+x_{{0}} \right) }{x_{{0}}}^{{
\frac{2}{3}}}\sqrt [3]{x} \left( 3\,{b}^{2}+{x_{{0}}}^{2} \right) ^{-{
\frac{5}{6}}}{\frac {1}{\sqrt [6]{3\,{b}^{2}+{x}^{2}}}}}
\nonumber \\
-{\frac {2\,v_
{{0}}}{3}\sqrt [3]{1+\epsilon\, \left( -x+x_{{0}} \right) } \left( 3\,
{b}^{2}+{x}^{2} \right) ^{{\frac{5}{6}}}{x_{{0}}}^{{\frac{2}{3}}}
 \left( 3\,{b}^{2}+{x_{{0}}}^{2} \right) ^{-{\frac{5}{6}}}{x}^{-{\frac
{5}{3}}}}=0
\quad  .
\end{eqnarray}

\noindent The solution for $x$ of the above minimum determines
the jet's length, $x_j$,
\begin{equation}
x_j =
\frac
{
4\,{b}^{2}{\epsilon}^{2}+{\epsilon}^{2}{x_{{0}}}^{2}+\sqrt [3]{D_{{2}}
}\epsilon\,x_{{0}}+{D_{{2}}}^{{\frac{2}{3}}}+2\,\epsilon\,x_{{0}}+
\sqrt [3]{D_{{2}}}+1
}
{
4\,\epsilon\,\sqrt [3]{D_{{2}}}
}
\quad ,
\end{equation}
where
\begin{eqnarray}
D_1=
-16\,{b}^{4}{\epsilon}^{4}+429\,{b}^{2}{\epsilon}^{4}{x_{{0}}}^{2}-24
\,{\epsilon}^{4}{x_{{0}}}^{4}+858\,{b}^{2}{\epsilon}^{3}x_{{0}}
\nonumber\\
-96\,{
\epsilon}^{3}{x_{{0}}}^{3}
+429\,{b}^{2}{\epsilon}^{2}-144\,{\epsilon}^
{2}{x_{{0}}}^{2}-96\,\epsilon\,x_{{0}}-24
\quad ,
\end{eqnarray}
and
\begin{equation}
D_2=
-42\,{b}^{2}{\epsilon}^{3}x_{{0}}+{\epsilon}^{3}{x_{{0}}}^{3}-42\,{b}^
{2}{\epsilon}^{2}
+3\,{\epsilon}^{2}{x_{{0}}}^{2}+2\,b\sqrt {{\it D1}}
\epsilon+3\,\epsilon\,x_{{0}}+1
\quad  .
\end{equation}

\begin{figure*}
\begin{center}
\includegraphics[width=6.6cm]{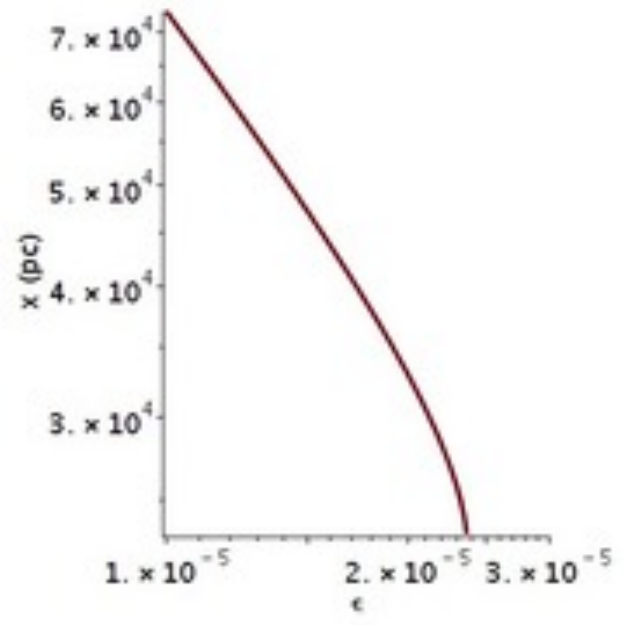}
\end {center}
\caption
{
Length  of the  jet
for a   Lane--Emden profile,
$x_j$,  in pc   as a function of $\epsilon$,
with $b$ as  in Table \ref{jetparameters}.
}
\label{xlunpc}
    \end{figure*}

\noindent Figure \ref{xlunpc} shows $x_j$  numerically.
%figure xlunpc
% end figure xlunpc

\section{Relativistic Turbulent Jets}
\label{secrelativistic}

The conservation of the   energy flux in special relativity (SR)
in  the presence of a  velocity $v$ along one direction
states that
\begin{equation}
A(x) \frac { 1}{ 1 -\frac {v^2}{c^2}} (e_0 +p_0) v = cost
\label{enthalpy}
\end{equation}

\noindent where

\begin {equation}
A(x)=\pi ( x   \tan ( \frac{\alpha}{2}))^2
\label{ax}
\quad ,
\end{equation}
\\
\noindent  is the considered area in the direction perpendicular
to the motion,
$c$ is the speed of light,
$e_0= c^2 \rho$ is the energy density in the rest
frame of the moving fluid,
and $p_0$ is the pressure in the rest frame
of the moving fluid,
see  formula A31 in
\cite{deyoung}.
In accordance with the current models of classical turbulent jets,
we  insert $p_0=0$  and
the   conservation law
for relativistic energy flux
is
\begin{equation}
\rho  c^2 v \frac { 1}{ 1 -\frac {v^2}{c^2} } A(x) = cost
\quad .
\label{relativisticflux}
\end{equation}
\index{conservation~energy~flux!relativistic}
\noindent Our physical units are
pc for length  and  yr for time, and
in these units, the speed of light is
$c=0.306$  \ pc \ yr$^{-1}$.
A discussion of the mass--energy equivalence principle in fluids
can be found in \cite{Palacios2015a}.

\subsection{Constant Density in SR}

The  conservation of the relativistic energy flux
when the density is constant can be written
as a differential equation
\begin{eqnarray}
{\rho\,{c}^{2} \left( {\frac {\rm d}{{\rm d}t}}x \left( t \right)
 \right) \pi \, \left( x \left( t \right)  \right) ^{2} \left( \tan
 \left( \frac{\alpha}{2} \right)  \right) ^{2} \left( 1-{\frac { \left( {\frac
{\rm d}{{\rm d}t}}x \left( t \right)  \right) ^{2}}{{c}^{2}}}
 \right) ^{-1}}
 \nonumber \\
 -{\rho\,{c}^{2}v_{{0}}\pi \,{x_{{0}}}^{2} \left( \tan
 \left( \frac{\alpha}{2} \right)  \right) ^{2} \left(1 -{\frac {{v_{{0}}}^{2}}{{c
}^{2}}} \right) ^{-1}}=0
\label{eqndiffrel}
\quad .
\end{eqnarray}
\\\\
\noindent Although an analytical solution of the previous differential equation
at the moment of writing does not
exist, we can
provide a  power series solution of the form\\
\begin{equation}
x(t) = a_0 +a_1  t +a_2 t^2 +a_3  t^3 + \dots
\label{xtrelseries}
\end{equation}
\\
\noindent see  \cite{Tenenbaum1963,Ince2012}.
The coefficients $a_n$ up to order 4  are
\begin{eqnarray}
a_0=&  x_{{0}}      \nonumber \\
a_1=&  v_{{0}}    \nonumber  \\
a_2=&  \frac{1}{3}\,{\frac {{v_{{0}}}^{3} \left( 5\,{c}^{6}-11\,{c}^{4}{v_{{0}}}^{2}+
3\,{c}^{2}{v_{{0}}}^{4}+3\,{v_{{0}}}^{6} \right) }{{x_{{0}}}^{2}
 \left( {c}^{2}+{v_{{0}}}^{2} \right)  \left( {c}^{4}+2\,{c}^{2}{v_{{0
}}}^{2}+{v_{{0}}}^{4} \right) }}
    \nonumber \\
a_3=&
\frac{1}{3}\,{\frac {{v_{{0}}}^{3} \left( 5\,{c}^{6}-11\,{c}^{4}{v_{{0}}}^{2}+
3\,{c}^{2}{v_{{0}}}^{4}+3\,{v_{{0}}}^{6} \right) }{{x_{{0}}}^{2}
 \left( {c}^{2}+{v_{{0}}}^{2} \right)  \left( {c}^{4}+2\,{c}^{2}{v_{{0
}}}^{2}+{v_{{0}}}^{4} \right) }}
\quad .
\end{eqnarray}

\begin{figure*}
\begin{center}
\includegraphics[width=7cm]{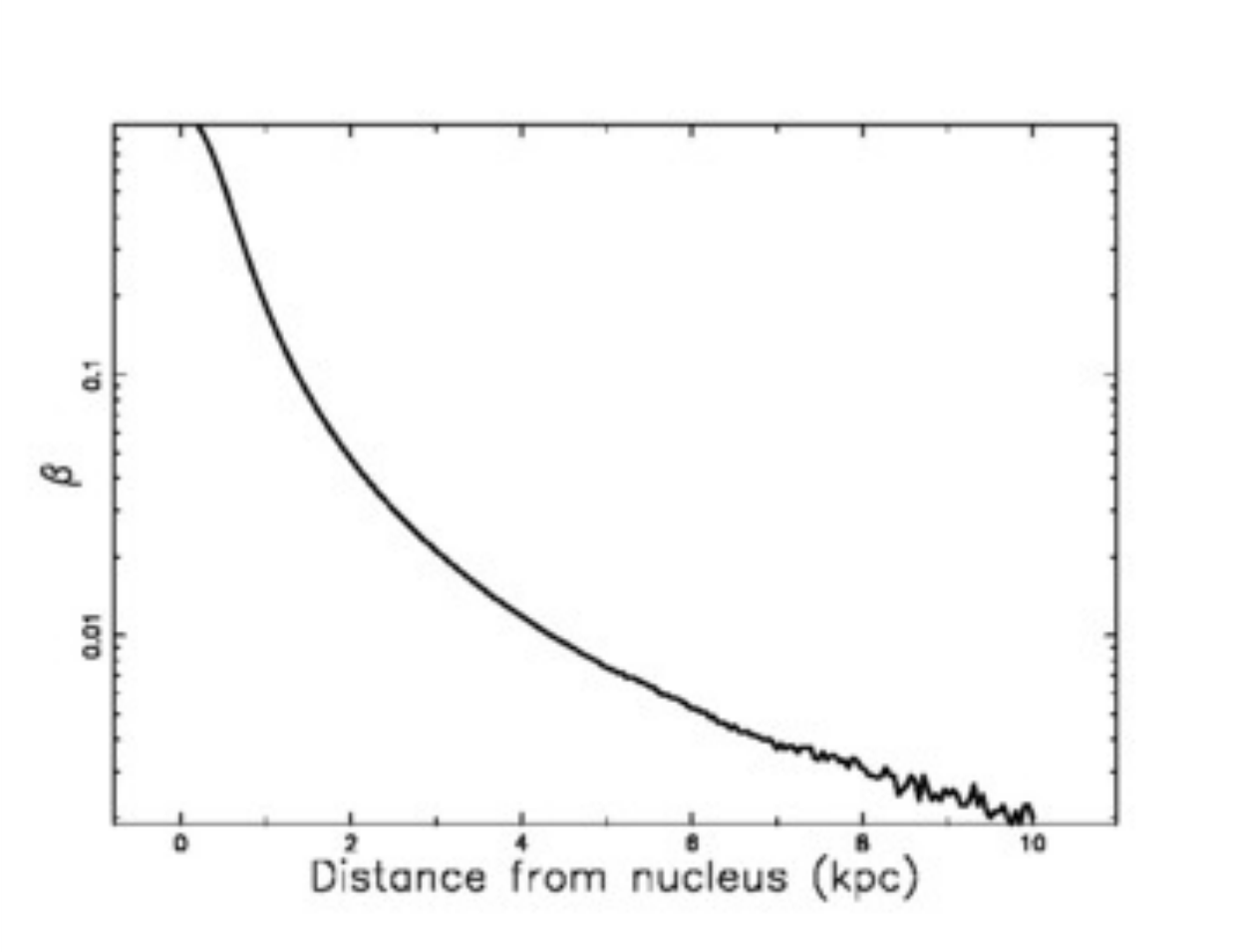}
\end {center}
\caption
{
Relativistic $\beta$
for constant  density
   as a function
of  the distance from the nucleus  when
$x_0$ =200~pc and $\beta_0$ =0.9 in the case of constant
density.
}
\label{energybeta}
    \end{figure*}

To find  a numerical solution of this
differential  equation,
we isolate the velocity from
Eq.~(\ref{eqndiffrel})
\begin{eqnarray}
v(x;x_0,\beta_0,c) = \nonumber \\
\frac{1}{2}\,{\frac { \left( {\beta_{{0}}}^{2}{x}^{2}-{x}^{2}+\sqrt {{x}^{4}{
\beta_{{0}}}^{4}-2\,{x}^{4}{\beta_{{0}}}^{2}+4\,{\beta_{{0}}}^{2}{x_{{0
}}}^{4}+{x}^{4}} \right) c}{\beta_{{0}}{x_{{0}}}^{2}}}
\,
\label{vxrel}
\end{eqnarray}
where $\beta_0=\frac{v_0}{c}$
and  separate the variables
\begin{eqnarray}
\int_{x_0}^x
2\,{\frac {\beta_{{0}}{x_{{0}}}^{2}}{ \left( {\beta_{{0}}}^{2}{x}^{2}-
{x}^{2}+\sqrt {{x}^{4}{\beta_{{0}}}^{4}-2\,{x}^{4}{\beta_{{0}}}^{2}+4
\,{\beta_{{0}}}^{2}{x_{{0}}}^{4}+{x}^{4}} \right) c}}
dx  \nonumber \\= \int_0^t  dt
\quad .
\end{eqnarray}
The indefinite integral on the left side
of the previous equation
has an analytical expression
\begin{equation}
I(x;\beta_0,c,x_0)=
\frac{AN}{AD}
\end{equation}
where
\begin{eqnarray}
AN =& \nonumber \\
2\,{\beta_{{0}}}^{3}{x_{{0}}}^{6}\sqrt {2}\sqrt {4-2\,{\frac {i\beta_{
{0}}{x}^{2}}{{x_{{0}}}^{2}}}+2\,{\frac {i{x}^{2}}{\beta_{{0}}{x_{{0}}}
^{2}}}}\sqrt {4+2\,{\frac {i\beta_{{0}}{x}^{2}}{{x_{{0}}}^{2}}}-2\,{
\frac {i{x}^{2}}{\beta_{{0}}{x_{{0}}}^{2}}}}\times
~& \nonumber \\
 \times {\it F} ( 1/2
\,x\sqrt {2}\sqrt {{\frac {i ( {\beta_{{0}}}^{2}-1 ) }{
\beta_{{0}}{x_{{0}}}^{2}}}},i )
\nonumber \\ -{\beta_{{0}}}^{3}{x_{{0}}}^{2}{
x}^{3}\sqrt {{\frac {i\beta_{{0}}}{{x_{{0}}}^{2}}}-{\frac {i}{\beta_{{0
}}{x_{{0}}}^{2}}}}\sqrt {{x}^{4}{\beta_{{0}}}^{4}-2\,{x}^{4}{\beta_{{0
}}}^{2}+4\,{\beta_{{0}}}^{2}{x_{{0}}}^{4}+{x}^{4}}
~&\nonumber \\
+\beta_{{0}}{x_{{0}}
}^{2}{x}^{3}\sqrt {{\frac {i\beta_{{0}}}{{x_{{0}}}^{2}}}-{\frac {i}{
\beta_{{0}}{x_{{0}}}^{2}}}}\sqrt {{x}^{4}{\beta_{{0}}}^{4}-2\,{x}^{4}{
\beta_{{0}}}^{2}+4\,{\beta_{{0}}}^{2}{x_{{0}}}^{4}+{x}^{4}}
&\nonumber \\
+{\beta_{{0
}}}^{5}{x_{{0}}}^{2}{x}^{5}\sqrt {{\frac {i\beta_{{0}}}{{x_{{0}}}^{2}}
}-{\frac {i}{\beta_{{0}}{x_{{0}}}^{2}}}}
-2\,{\beta_{{0}}}^{3}{x_{{0}}}
^{2}{x}^{5}\sqrt {{\frac {i\beta_{{0}}}{{x_{{0}}}^{2}}}-{\frac {i}{
\beta_{{0}}{x_{{0}}}^{2}}}}
~& \nonumber \\
+4\,{\beta_{{0}}}^{3}{x_{{0}}}^{6}x\sqrt {{
\frac {i\beta_{{0}}}{{x_{{0}}}^{2}}}-{\frac {i}{\beta_{{0}}{x_{{0}}}^{
2}}}}+\beta_{{0}}{x_{{0}}}^{2}{x}^{5}\sqrt {{\frac {i\beta_{{0}}}{{x_{
{0}}}^{2}}}-{\frac {i}{\beta_{{0}}{x_{{0}}}^{2}}}}
\end{eqnarray}
and
\begin{eqnarray}
AD= \nonumber \\
6\,c{\beta_{{0}}}^{2}{x_{{0}}}^{4}\sqrt {{\frac {i\beta_{{0}}}{{x_{{0}
}}^{2}}}-{\frac {i}{\beta_{{0}}{x_{{0}}}^{2}}}}\sqrt {{x}^{4}{\beta_{{0
}}}^{4}-2\,{x}^{4}{\beta_{{0}}}^{2}+4\,{\beta_{{0}}}^{2}{x_{{0}}}^{4}+
{x}^{4}}
\end{eqnarray}
where $i=\sqrt{-1}$  and
\begin{equation}
F(x;m)=\int_{0}^{x}\!{\frac {1}{\sqrt {1-{{\it t}}^{2}}\sqrt {1-{m}^{2
}{{\it t}}^{2}}}}\,{\rm d}{\it t}
\end{equation}
is the elliptic integral of the first kind,
see formula 17.2.7 in \cite{Abramowitz1965}.
Figure \ref{energybeta} shows the
%citiamofigura_energybeta
behaviour of $\beta$ as  function
of the distance.
% figure   energybeta
% end energybeta

\begin{figure*}
\begin{center}
\includegraphics[width=8cm]{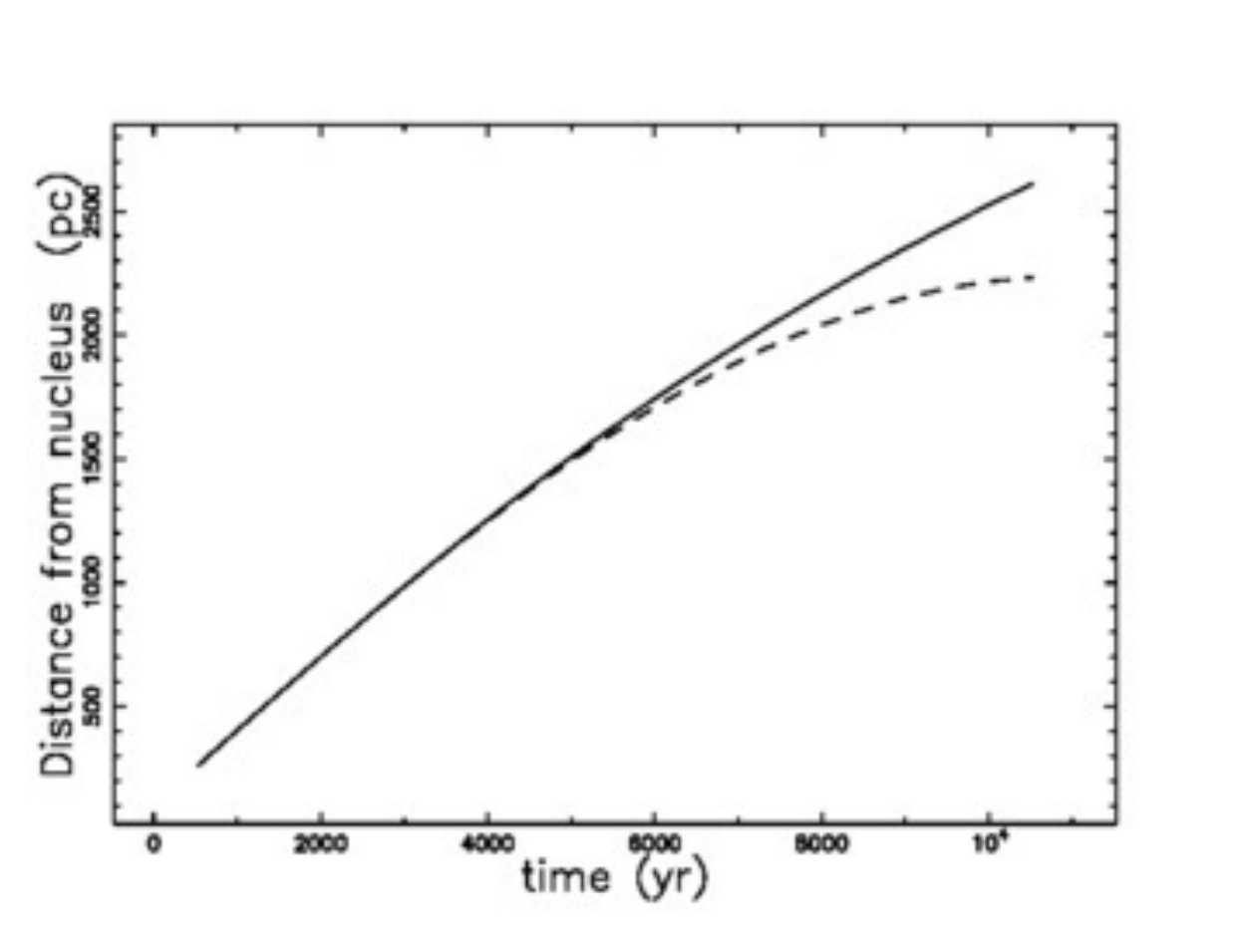}
\end {center}
\caption
{
Non-linear relativistic solution
for constant  density
 as given
by Eq. (\ref{solrelativisticnl}) (full line)
and series solution
 as given
by Eq. (\ref{xtrelseries}) (dashed line)
when
$x_0$ =100 pc and $\beta_0$ =0.999.
}
\label{energyrelxt}
    \end{figure*}

A numerical  solution can be found  by solving
the following non-linear  equation
\begin{equation}
I(x;\beta_0,c,x_0)- I(x_0;\beta_0,c,x_0)=t
\label{solrelativisticnl}
\end{equation}
 and  Figure  \ref{energyrelxt}
%citiamofigura_energyrelxt
presents a typical comparison with the series solution.
%figure energyrelxt
% end figure energyrelxt
The relativistic rate of mass flow in the case of
constant density is
{\small
\begin{eqnarray}
\dot {m}(x) =  & ~ \nonumber \\
\frac
{
\rho\, \left( {\beta_{{0}}}^{2}{x}^{2}-{x}^{2}+\sqrt {{x}^{4}{\beta_{{0
}}}^{4}-2\,{x}^{4}{\beta_{{0}}}^{2}+4\,{\beta_{{0}}}^{2}{x_{{0}}}^{4}+
{x}^{4}} \right) c\pi \,x \left( \tan \left( \alpha/2 \right)
 \right) ^{2}
}
{
\sqrt {2 \left( 1-{\beta_{{0}}}^{2} \right)  \left( {\beta_{{0}}}^{
2}{x}^{2}-{x}^{2}+\sqrt {{x}^{4}{\beta_{{0}}}^{4}-2\,{x}^{4}{\beta_{{0
}}}^{2}+4\,{\beta_{{0}}}^{2}{x_{{0}}}^{4}+{x}^{4}} \right) }
}
& ~
\end{eqnarray}
}
%siamoqui

\subsection{Inverse Power Law  Profile of Density in SR }

The  conservation of the relativistic energy  flux
in the presence  of an inverse power law
density profile
as given by Eq. (\ref{profpower})
is
\begin{eqnarray}
{\rho_0\,{c}^{2} \left( {\frac {\rm d}{{\rm d}t}}x \left( t \right)
 \right) \pi \, \left( x \left( t \right)  \right) ^{2} \left( \tan
 \left( \frac{\alpha}{2} \right)  \right) ^{2} \left( {\frac {x_{{0}}}{x
 \left( t \right) }} \right) ^{\delta} \left( -{\frac { \left( {\frac
{\rm d}{{\rm d}t}}x \left( t \right)  \right) ^{2}}{{c}^{2}}}+1
 \right) ^{-1}} \nonumber \\
-{\rho_0\,{c}^{2}v_{{0}}\pi \,{x_{{0}}}^{2} \left( \tan
 \left( \frac{\alpha}{2} \right)  \right) ^{2} \left( -{\frac {{v_{{0}}}^{2}}{
{c}^{2}}}+1 \right) ^{-1}}=0
\quad.
\label{diffeqnrelpower}
\end{eqnarray}
This  differential equation does not have an analytical
solution. An expression for
$\beta$ as a function of the distance
is
\begin{equation}
\beta(x) =
\frac{1}{2}\,{\frac {1}{\beta_{{0}}{x_{{0}}}^{2}} \left( {\beta_{{0}}}^{2}{x}^
{2} \left( {\frac {x_{{0}}}{x}} \right) ^{\delta}-{x}^{2} \left( {
\frac {x_{{0}}}{x}} \right) ^{\delta}+\sqrt {D} \right) }
\label{betadistance}
\end{equation}
with
\begin{eqnarray}
D= &
\left(  \left( {\frac {x_{{0}}}{x}} \right) ^{\delta} \right) ^{2}{
\beta_{{0}}}^{4}{x}^{4}-2\, \left(  \left( {\frac {x_{{0}}}{x}}
 \right) ^{\delta} \right) ^{2}{\beta_{{0}}}^{2}{x}^{4}+ \left(
 \left( {\frac {x_{{0}}}{x}} \right) ^{\delta} \right) ^{2}{x}^{4}+4\,
{\beta_{{0}}}^{2}{x_{{0}}}^{4}.
\label{eqnd}
\end{eqnarray}

\begin{figure*}
\begin{center}
\includegraphics[width=8cm]{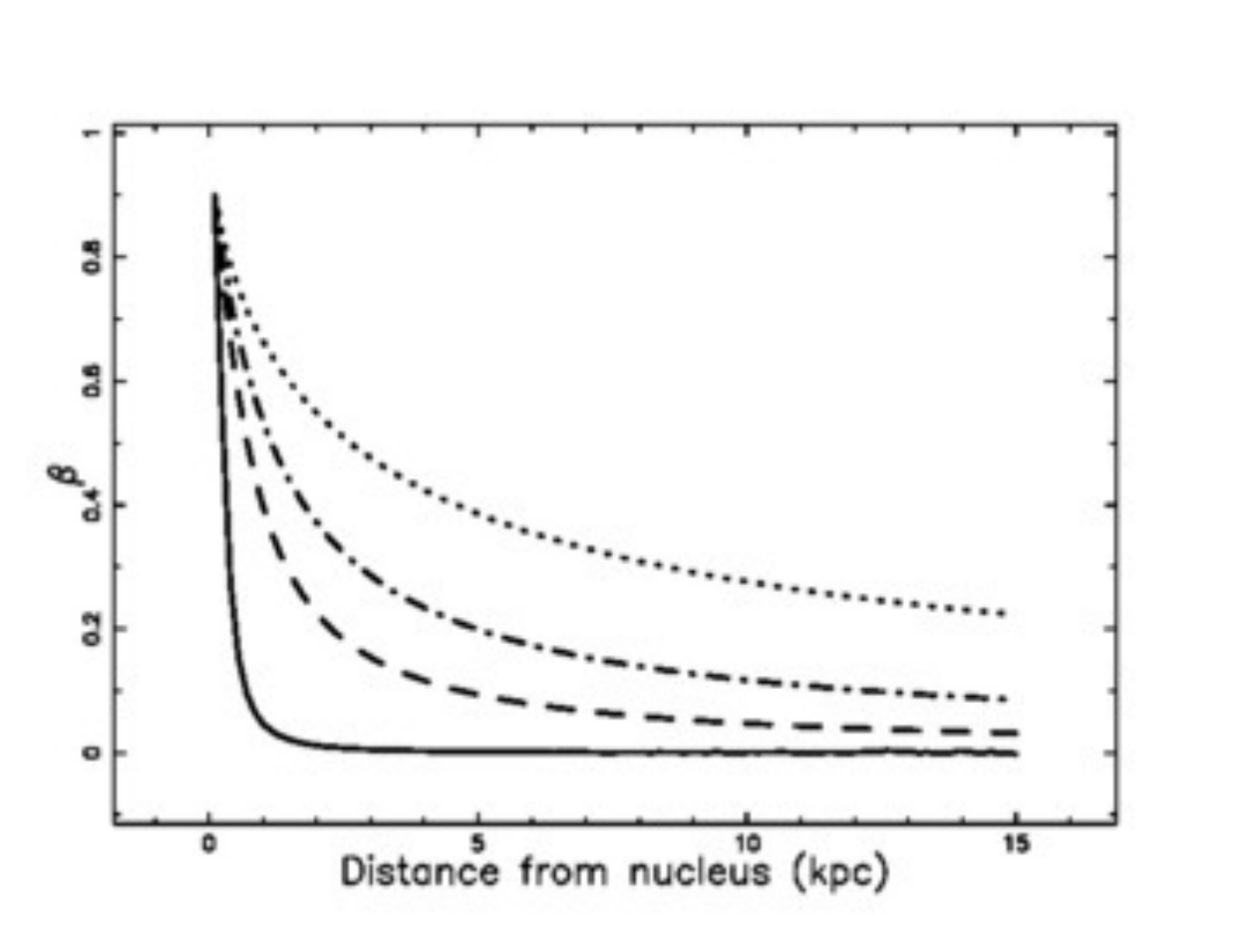}
\end {center}
\caption
{
Relativistic $\beta$ for the relativistic
energy flux conservation
in the presence  of an inverse power law
  as a function
of  the distance from the nucleus  when
$x_0$ =100~pc and $\beta_0$ =0.9:
$\delta =0$   (full line),
$\delta =1$   (dashes),
$\delta =1.2$ (dot-dash-dot-dash)
and
$\delta =1.4$ (dotted).
}
\label{betaxdeltaenergy}
    \end{figure*}

\begin{figure*}
\begin{center}
\includegraphics[width=7cm]{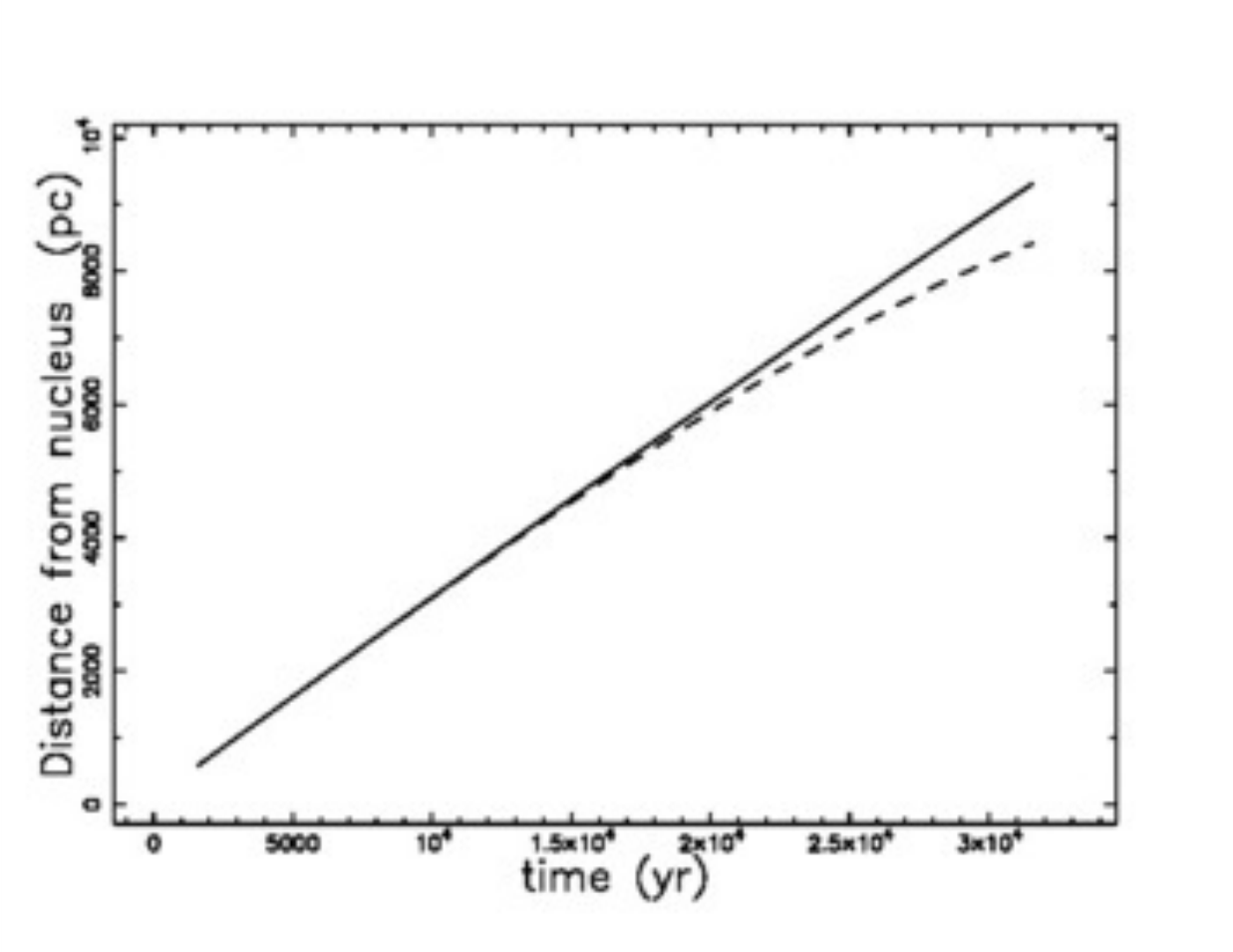}
\end {center}
\caption
{
Non-linear relativistic solution
in presence  of an inverse power law
as given
by Eq. (\ref{diffeqnrelpower}) (full line)
and series solution
 as given
by Eq. (\ref{xtseriesreldelta}) (dashed line)
when
$x_0$ =100 pc and $\beta_0$ =0.999.
}
\label{energyrelxtdelta}
    \end{figure*}

The behaviour of $\beta$ as a function of the distance
for different values of $\delta$ can be seen
in Figure \ref{betaxdeltaenergy}.
%citiamofigura_betaxdeltaenergy
% figure   betaxdeltaenergy
% end betaxdeltaenergy
A  power series solution  for this differential
equation (\ref{diffeqnrelpower})
up to order three  gives
\begin{eqnarray}
a_0=& x_{{0}}       \nonumber \\
a_1=& v_{{0}}      \nonumber  \\
a_2=& \frac{1}{2}\,{\frac {{v_{{0}}}^{2} \left( {c}^{2}\delta-\delta\,{v_{{0}}}^{2}-
2\,{c}^{2}+2\,{v_{{0}}}^{2} \right) }{x_{{0}} \left( {c}^{2}+{v_{{0}}}
^{2} \right) }}
\label{xtseriesreldelta}
\quad .
\end{eqnarray}
Figure  \ref{energyrelxtdelta}
%citiamofigura_energyrelxtdelta
shows a comparison between the numerical solution
of (\ref{diffeqnrelpower})
with the series solution.
%figure energyrelxtdelta
% end figure energyrelxtdelta
\newpage
The relativistic rate of mass flow in the case of an inverse power law for
the density is
\begin{equation}
\dot {m}(x) =
\frac
{
\rho_{{0}} \left( {\frac {x_{{0}}}{x}} \right) ^{\delta} \left( {\beta
_{{0}}}^{2}{x}^{2} \left( {\frac {x_{{0}}}{x}} \right) ^{\delta}-{x}^{
2} \left( {\frac {x_{{0}}}{x}} \right) ^{\delta}+\sqrt {D} \right) c
\pi \,{x}^{2} \left( \tan \left( \alpha/2 \right)  \right) ^{2}
}
{
2\,\beta_{{0}}{x_{{0}}}^{2}\sqrt {-1/4\,{\frac {1}{{\beta_{{0}}}^{2}{x
_{{0}}}^{4}} \left( {\beta_{{0}}}^{2}{x}^{2} \left( {\frac {x_{{0}}}{x
}} \right) ^{\delta}-{x}^{2} \left( {\frac {x_{{0}}}{x}} \right) ^{
\delta}+\sqrt {D} \right) ^{2}}+1}
}
\end{equation}
where  $\rho_0$ is the density at $x_0$ and
$D$ was defined in Eq.~(\ref{eqnd}).

\subsection{The Lane--Emden Profile}

\begin{figure*}
\begin{center}
\includegraphics[width=7cm]{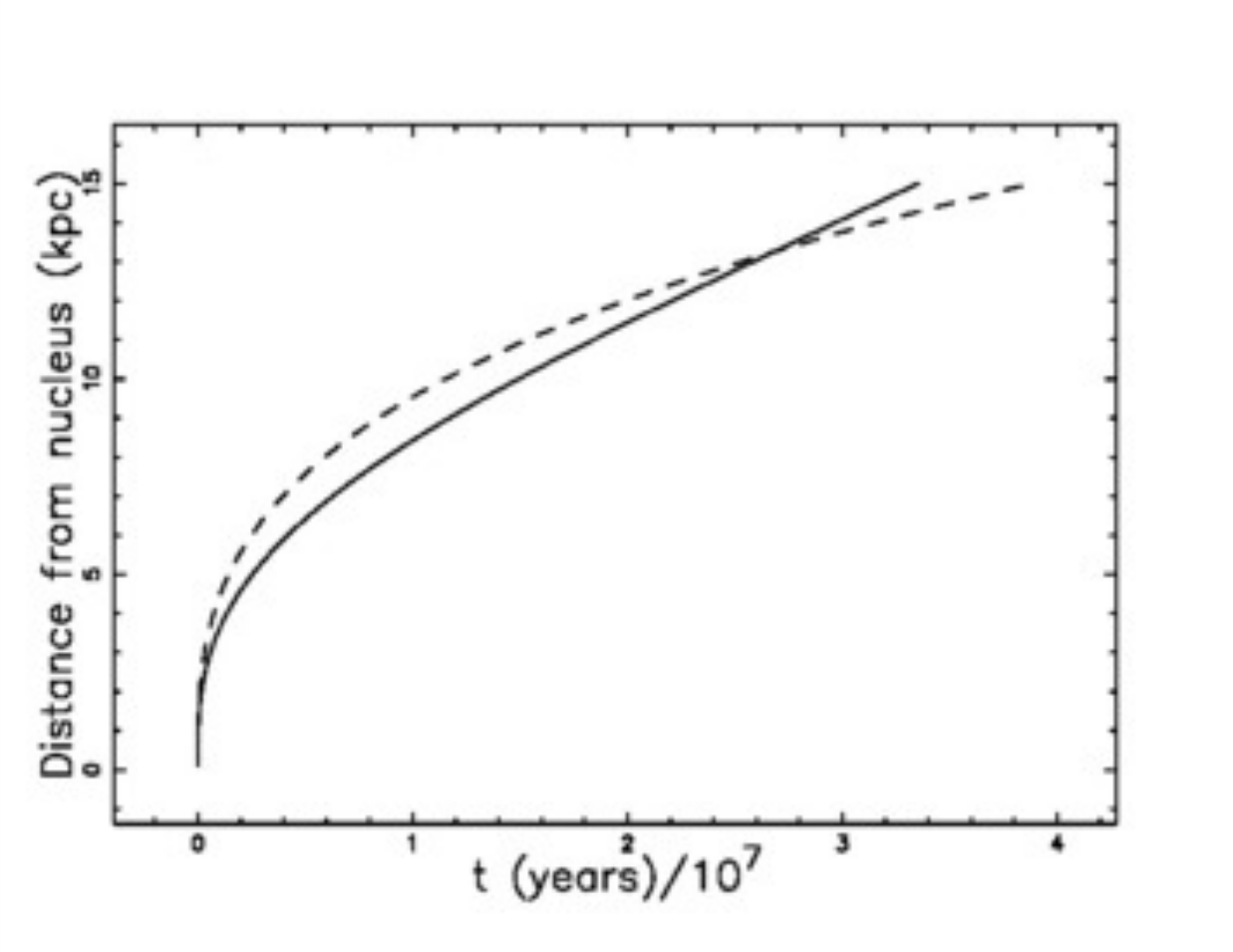}
\end {center}
\caption
{
Numerical relativistic solution
for a Lane--Emden  ($n=5$) density profile
as given
by equation (\ref{eqnmotionrel}) (full line)
and truncated series expansion
as given
by equation (\ref{traj_rel_series}) (dashed line),
with parameters as in Table \ref{jetrel}.
}
\label{traj_rel_series}
    \end{figure*}

In the presence of a
Lane--Emden  ($n=5$) density profile, as given
by equation (\ref{densita5b}) and $A(x)$ as given
by equation~(\ref{ax}),
the conservation
of     relativistic   flux of energy for
a straight jet
takes the form
\begin{equation}
\frac
{
\rho_{{0}}{c}^{3}\beta\,\pi\,{x}^{2} \left( \tan \left( \alpha/2
 \right)  \right) ^{2}
}
{
\left( 1+\frac{1}{3}\,{\frac {{x}^{2}}{{b}^{2}}} \right) ^{5/2} \left(1- {\beta
}^{2} \right)
}
=
\frac
{
\rho_{{0}}{c}^{3}\beta0\,\pi\,{x_{{0}}}^{2} \left( \tan \left( \alpha
/2 \right)  \right) ^{2}
}
{
 \left( 1+\frac{1}{3}\,{\frac {{x_{{0}}}^{2}}{{b}^{2}}} \right) ^{5/2} \left(
{1-\beta0}^{2} \right)
}
 ,
\end{equation}
where
$v$ is the velocity at $x$,
$v_0$ is the velocity at $x_0$,
$\beta=\frac{v}{c}$ and $\beta_0=\frac{v_0}{c}$.
The solution for $\beta$ to first order is
\begin{equation}
\beta(x;x_0,b,\beta_0)=\frac
{N}
{
 \left( 1+\frac{1}{3}\,{\frac {{x_{{0}}}^{2}}{{b}^{2}}} \right) ^{5/2} \left(
{\beta0}^{2}-1 \right)
}
 ,
\label{betarelfirst}
\end{equation}
where
{\footnotesize
\begin{eqnarray}
N=
9\,\sqrt {3\,{b}^{2}+{x_{{0}}}^{2}}{x}^{2}{b}^{4}{\beta_{{0}}}^{2}+6\,
\sqrt {3\,{b}^{2}+{x_{{0}}}^{2}}{x}^{2}{b}^{2}{x_{{0}}}^{2}{\beta_{{0}
}}^{2}+\sqrt {3\,{b}^{2}+{x_{{0}}}^{2}}{x}^{2}{x_{{0}}}^{4}{\beta_{{0}
}}^{2}
\nonumber \\
-9\,\sqrt {3\,{b}^{2}+{x_{{0}}}^{2}}{x}^{2}{b}^{4}-6\,\sqrt {3\,
{b}^{2}+{x_{{0}}}^{2}}{x}^{2}{b}^{2}{x_{{0}}}^{2}-\sqrt {3\,{b}^{2}+{x
_{{0}}}^{2}}{x}^{2}{x_{{0}}}^{4}
\nonumber \\
+\Biggl (243\,{x}^{4}   ( {b}^{2}+ \frac{1}{3}
\,{x_{{0}}}^{2}   ) ^{5}{\beta_{{0}}}^{4}+   ( -2\,{x}^{4}{x_{
{0}}}^{10}-30\,{b}^{2}{x}^{4}{x_{{0}}}^{8}-180\,{b}^{4}{x}^{4}{x_{{0}}
}^{6}
\nonumber \\
+   ( 972\,{b}^{10}+1620\,{b}^{8}{x}^{2}+540\,{b}^{6}{x}^{4}+
360\,{b}^{4}{x}^{6}+60\,{b}^{2}{x}^{8}+4\,{x}^{10}   ) {x_{{0}}}^{
4}
\nonumber \\
-810\,{b}^{8}{x}^{4}{x_{{0}}}^{2}-486\,{b}^{10}{x}^{4}   ) {
\beta_{{0}}}^{2}+243\,{x}^{4}   ( {b}^{2}+\frac{1}{3}\,{x_{{0}}}^{2}
   ) ^{5}\Biggr )^{1/2}.
\end{eqnarray}
}
The equation for the relativistic trajectory   is
\begin{equation}
\int_{x_0}^x \frac{1}{\beta(x;x_0,b,\beta_0) \, c} dx  = t
\quad .
\label{eqnmotionrel}
\end{equation}
The integral in this equation does not have an analytical
solution
and should be integrated numerically.
To have analytical results,
two approximations are now introduced.
The first approximation  computes a truncated series expansion
for the integrand of the integral in equation (\ref{eqnmotionrel}),
which transforms the relativistic equation of motion
into
\begin{equation}
F(x) - F(x_0) = t
\label{eqn_traj_rel_series}
\quad ,
\end{equation}
with
\begin{equation}
F(x) = \frac{NF}{162\,{x_{{0}}}^{2}\beta_{{0}}{b}^{10}c}
\quad ,
\end{equation}
where
{\footnotesize
\begin{eqnarray}
NF =  \left( {b}^{2} \right) ^{5/2}x \Bigl( 9\,\sqrt {3}\sqrt {3\,{b}^{2}+
{x_{{0}}}^{2}}{b}^{4}{\beta_{{0}}}^{2}{x}^{2}+6\,\sqrt {3}\sqrt {3\,{b
}^{2}+{x_{{0}}}^{2}}{b}^{2}{\beta_{{0}}}^{2}{x}^{2}{x_{{0}}}^{2}
\nonumber \\
+
\sqrt {3}\sqrt {3\,{b}^{2}+{x_{{0}}}^{2}}{\beta_{{0}}}^{2}{x}^{2}{x_{{0
}}}^{4}-9\,\sqrt {3}\sqrt {3\,{b}^{2}+{x_{{0}}}^{2}}{b}^{4}{x}^{2}
\nonumber  \\
-6\,
\sqrt {3}\sqrt {3\,{b}^{2}+{x_{{0}}}^{2}}{b}^{2}{x}^{2}{x_{{0}}}^{2}-
\sqrt {3}\sqrt {3\,{b}^{2}+{x_{{0}}}^{2}}{x}^{2}{x_{{0}}}^{4}-162\,
\sqrt {{b}^{10}{x_{{0}}}^{4}{\beta_{{0}}}^{2}} \Bigr )
.
\end{eqnarray}
}
\noindent\noindent  In this analytical  result we have time as
a function of the distance,
see  Figure~\ref{traj_rel_series}
where the percentage error at $x=15$\ kpc is
$\delta= 15.91\%$.

%figure traj_rel_series
% end figure traj_rel_series

%begin table jetrel
\begin{table}[ht!]
\caption
{
Parameters for a relativistic extra-galactic jet.
}
\label{jetrel}
\begin{center}
\begin{tabular}{|c|c|}
\hline
parameter    &  value    \\
\hline
$x_0$ (pc)   & 100       \\
$\beta_0$    & 0.9       \\
$b$   (pc)   & 10000\\
\hline
\end{tabular}
\end{center}
\end{table}
%end  table jetrel

\begin{figure*}
\begin{center}
\includegraphics[width=7cm]{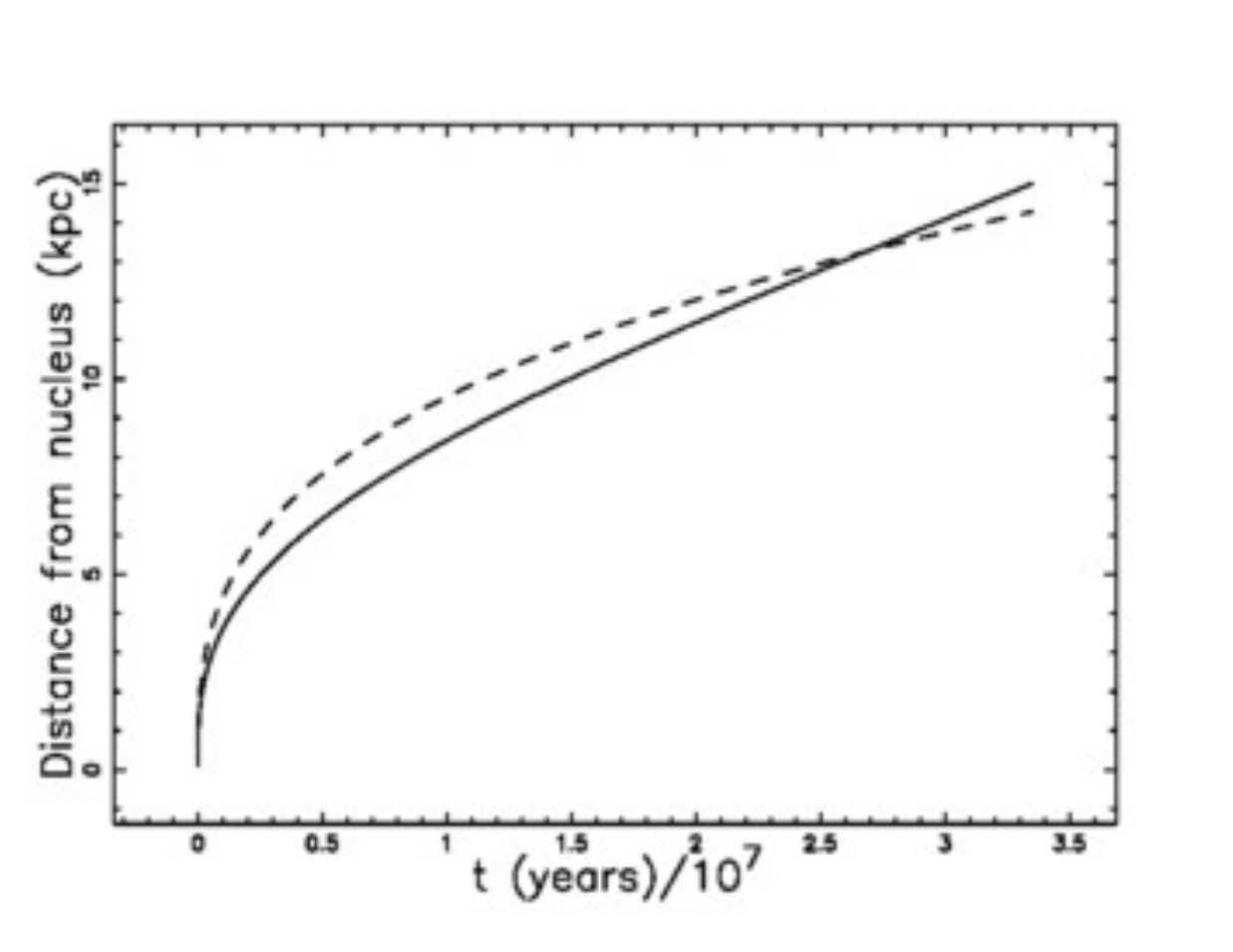}
\end {center}
\caption
{
Numerical relativistic solution
for a Lane--Emden  ($n=5$) density profile
as given
by equation (\ref{eqnmotionrel}) (full line)
and Pad\'e approximant
 as given
by equation (\ref{traj_rel_pade}) (dashed line),
with parameters as in Table \ref{jetrel}.
}
\label{traj_rel_pade}
    \end{figure*}

The second approximation  computes a
Pad\'e  approximant of order  [2/1],
see \cite{2012Adachi,Aviles2014,Wei2014},
for the integrand of the integral in equation (\ref{eqnmotionrel})
\begin{equation}
P(x) - P(x_0) = t
\quad ,
\end{equation}
with
\begin{equation}
P(x) \frac{NP}{162\,{b}^{10}{x_{{0}}}^{4}{\beta_{{0}}}^{2}c}
\quad ,
\end{equation}
where
{\small
\begin{eqnarray}
&NP(x) = -x \left( {b}^{2} \right) ^{5/2}{x_{{0}}}^{2}\beta_{{0}} \Big ( 9\,
\sqrt {3}\sqrt {3\,{b}^{2}+{x_{{0}}}^{2}}{b}^{4}{\beta_{{0}}}^{2}{x}^{
2}
\nonumber  \\
&+6\,\sqrt {3}\sqrt {3\,{b}^{2}+{x_{{0}}}^{2}}{b}^{2}{\beta_{{0}}}^{2
}{x}^{2}{x_{{0}}}^{2}+\sqrt {3}\sqrt {3\,{b}^{2}+{x_{{0}}}^{2}}{\beta_
{{0}}}^{2}{x}^{2}{x_{{0}}}^{4}
\nonumber \\
&-9\,\sqrt {3}\sqrt {3\,{b}^{2}+{x_{{0}}}
^{2}}{b}^{4}{x}^{2}-6\,\sqrt {3}\sqrt {3\,{b}^{2}+{x_{{0}}}^{2}}{b}^{2
}{x}^{2}{x_{{0}}}^{2}-\sqrt {3}\sqrt {3\,{b}^{2}+{x_{{0}}}^{2}}{x}^{2}
{x_{{0}}}^{4}
\nonumber \\
&-162\,\sqrt {{b}^{10}{x_{{0}}}^{4}{\beta_{{0}}}^{2}}
 \Big )
 .
\end{eqnarray}
}
\\
\noindent \noindent Although this equation can be inverted,  the
analytical expression for $x=G(t;x_0,\beta_0,b)$ as a function of time
is complicated
and is consequently omitted here.
As an example, with the parameters of Table \ref{jetrel},
we have
\begin{equation}
G(t) =\frac{NG}{DG}
\label{eqn_traj_rel_pade}
\quad ,
\end{equation}
with
\begin{eqnarray}
NG =
-{ 2.9237\times 10^{-17}}\, \Bigl ( -{ 1.7397\times 10^{54}}
\,t-{ 5.8851\times 10^{56}}
\nonumber \\
+{ 2.9816\times 10^{20}}\,\sqrt {
{ 1.9201\times 10^{73}}+{ 2.3032\times 10^{70}}\,t
}
\nonumber  \\
{
+{
 3.4042\times 10^{67}}\,{t}^{2}} \Bigr ) ^{2/3}+{ 3.2399
\times 10^{21}}
\quad ,
\end{eqnarray}
and
{\small
\begin{eqnarray}
DG=
\Bigg (
-{ 1.7397\times 10^{54}}\,t-{ 5.8851\times 10^{56}
}
\nonumber \\
+{ 2.9816\times 10^{20}}\,\sqrt {{ 1.9201\times 10^{73}}+{
 2.3032\times 10^{70}}\,t+{ 3.4042\times 10^{67}}\,{t}^{2}}
\Bigg )^{\frac{1}{3}}
 .
\end{eqnarray}
}
An example is shown in
Figure \ref{traj_rel_pade}, where the percentage error at $x=15$\ kpc is
$\delta= 4.81\%$.

%figure traj_rel_pade
% end figure traj_rel_pade

\subsection{Relativistic Solution to Second Order}

\label{secrelativisticlosses}
\begin{figure*}
\begin{center}
\includegraphics[width=7cm]{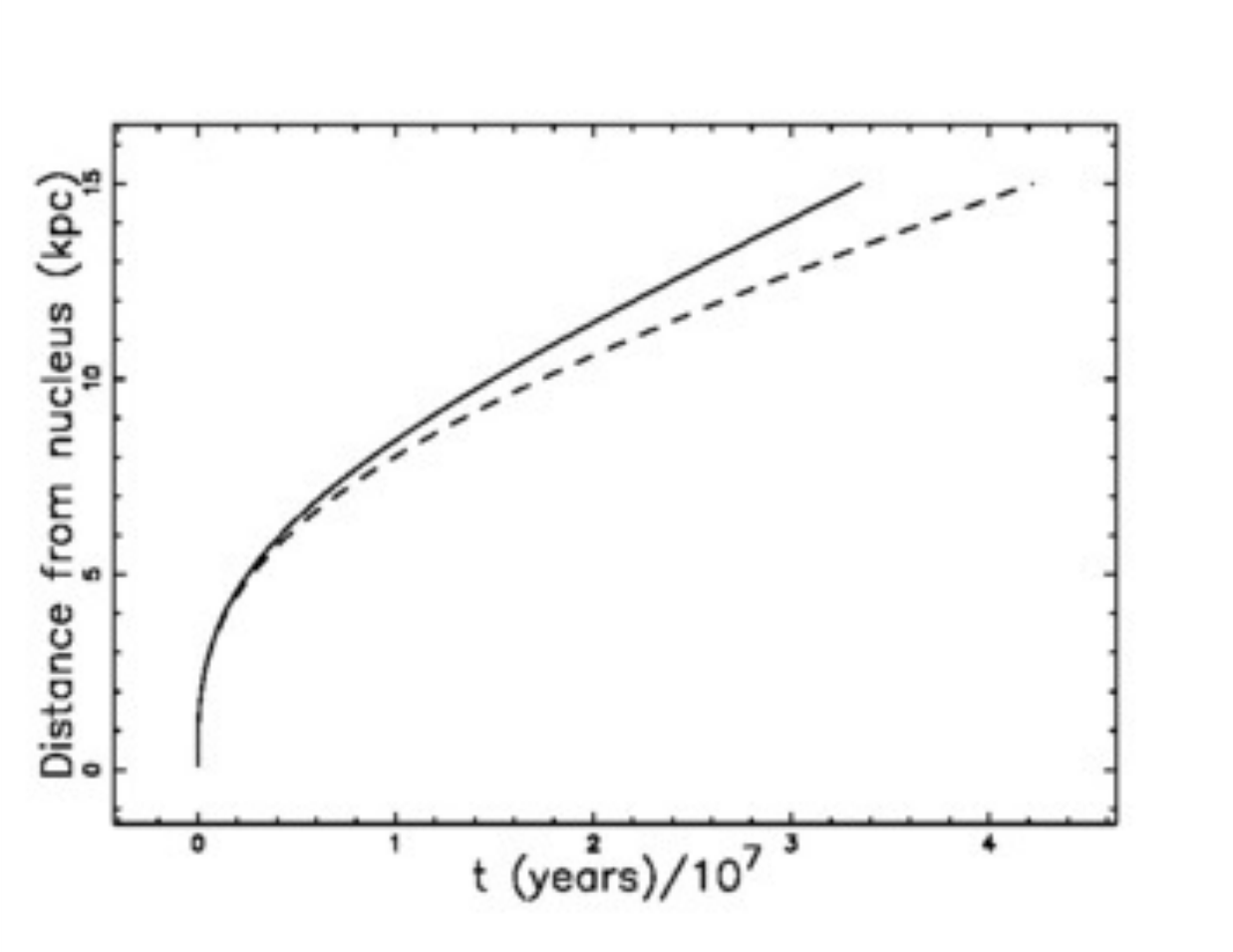}
\end {center}
\caption
{
Numerical relativistic solution
for a Lane--Emden  ($n=5$) density profile
as given
by equation (\ref{eqnmotionrel}) (full line)
and solution with back-reaction, i.e., to second order,
 (dashed line),
with parameters as in Table \ref{jetrel}
and $\epsilon=2.0\,10^{-5}$.
}
\label{traj_back_eps}
    \end{figure*}

\noindent \noindent We now  suppose that the radiative losses
for a Lane--Emden  ($n=5$) density profile
are  proportional to the    relativistic   flux of energy.
The integral of the losses, $L_r$, between $x_0$ and $x$ is
\begin{equation}
L_r(x;x_0,\beta_0,b,c) = - \epsilon
\frac
{
9\, \left( x-x_{{0}} \right) \rho0\,{c}^{3}\beta_{{0}}{x_{{0}}}^{2}
\pi\, \left( \tan \left( \alpha/2 \right)  \right) ^{2}{b}^{5}\sqrt {3
}
}
{
\left( 3\,{b}^{2}+{x_{{0}}}^{2} \right) ^{5/2} \left(1- {\beta_{{0}}}^{
2} \right)
}
\quad .
\label{relativisticlosses}
\end{equation}
The  conservation of the    relativistic   flux of energy in the presence
of  the back-reaction due to the radiative losses
is
\begin{eqnarray}
\frac{NR}
{
\left( 3\,{b}^{2}+{x}^{2} \right) ^{5/2} \left(
3\,{b}^{2}+{x_{{0}}}^
{2} \right) ^{5/2} \left( {\beta}^{2}-1 \right)  \left( {\beta_{{0}}}^
{2}-1 \right)
}=
\nonumber \\
\frac
{
9\,\rho0\,\sqrt {3}{c}^{3}\beta_{{0}}{x_{{0}}}^{2}{b}^{5}
}
{
 \left( 3\,{b}^{2}+{x_{{0}}}^{2} \right) ^{5/2} \left( {\beta_{{0}}}^{
2}-1 \right)
}
\quad ,
\end{eqnarray}
where
{\small
\begin{eqnarray}
&NR = 81\,\rho0\,{b}^{5}\sqrt {3} \Bigg  (    ( {b}^{2}+\frac{1}{3}\,{x}^{2}
   ) ^{2}\epsilon\,   ( \beta+1   ) \beta_{{0}}   (
\beta-1   )    ( x-x_{{0}}   ) {x_{{0}}}^{2}\sqrt {3\,{b}^{
2}+{x}^{2}}
\nonumber \\
&+   ( {b}^{2}+\frac{1}{3}\,{x_{{0}}}^{2}   ) ^{2}\beta\,{x}^
{2}   ( \beta_{{0}}+1   )    ( \beta_{{0}}-1   ) \sqrt {
3\,{b}^{2}+{x_{{0}}}^{2}} \Bigg  ) {c}^{3}.
\end{eqnarray}
}
The solution of this equation, to second order, for $\beta$
is
\begin{equation}
\beta= \frac{NB}
{
2\, \left( 3\,{b}^{2}+{x}^{2} \right) ^{5/2} \left(
\epsilon\,x-
\epsilon\,x_{{0}}-1 \right)  \left( 3\,{b}^{2}+{x_{{0}}}^{2} \right) {
x_{{0}}}^{2}\beta_{{0}}
}
\quad ,
\label{betarelsecond}
\end{equation}
where
\begin{eqnarray}
NB=
-\sqrt {3\,{b}^{2}+{x_{{0}}}^{2}}  \Bigg ( \sqrt {3\,{b}^{2}+{x_{{0}}}^{
2}} \times
\nonumber \\
\Big ( {{x}^{4}   ( \beta_{{0}}-1   ) ^{2}   ( \beta_{{0}}
+1   ) ^{2}{x_{{0}}}^{10}+15\,{b}^{2}{x}^{4}   ( \beta_{{0}}-1
   ) ^{2}   ( \beta_{{0}}+1   ) ^{2}{x_{{0}}}^{8}
}
\nonumber \\
{
+   (
972\,   ( {b}^{2}+\frac{1}{3}\,{x}^{2}   ) ^{5}{\beta_{{0}}}^{2}{
\epsilon}^{2}+90\,{b}^{4}{x}^{4}   ( \beta_{{0}}-1   ) ^{2}
   ( \beta_{{0}}+1   ) ^{2}   ) {x_{{0}}}^{6}
}
\nonumber \\
{
-1944\,
\epsilon\,   ( {b}^{2}+\frac{1}{3}\,{x}^{2}   ) ^{5}{\beta_{{0}}}^{2}
   ( \epsilon\,x-1   ) {x_{{0}}}^{5}+   ( 972\,   ( {b}^{
2}+\frac{1}{3}\,{x}^{2}   ) ^{5}{\beta_{{0}}}^{2}{x}^{2}{\epsilon}^{2}
}
\nonumber \\
{
-
1944\,   ( {b}^{2}+\frac{1}{3}\,{x}^{2}   ) ^{5}{\beta_{{0}}}^{2}x
\epsilon+4\,{x}^{10}{\beta_{{0}}}^{2}+60\,{b}^{2}{x}^{8}{\beta_{{0}}}^
{2}+360\,{b}^{4}{x}^{6}{\beta_{{0}}}^{2}
}
\nonumber \\
{
+270\,{b}^{6}   ( {\beta_{{0
}}}^{2}+1   ) ^{2}{x}^{4}+1620\,{b}^{8}{x}^{2}{\beta_{{0}}}^{2}+
972\,{b}^{10}{\beta_{{0}}}^{2}   ) {x_{{0}}}^{4}
}
\nonumber \\
{
+405\,{b}^{8}{x}^{
4}   ( \beta_{{0}}-1   ) ^{2}   ( \beta_{{0}}+1   ) ^{2}
{x_{{0}}}^{2}+243\,{b}^{10}{x}^{4}   ( \beta_{{0}}-1   ) ^{2}
   ( \beta_{{0}}+1   ) ^{2}}
\Big )^{1/2}
\nonumber \\
+27\,   ( {b}^{2}+\frac{1}{3}\,{x_{{0}}}
^{2}   ) ^{3}   ( \beta_{{0}}+1   ) {x}^{2}   ( \beta_{{0
}}-1   )   \Bigg )
\quad .
\end{eqnarray}
The relativistic equation of motion with back-reaction
can be solved by numerically integrating the  relation
in equation (\ref{eqnmotionrel}).
Figure \ref{traj_back_eps} gives an example.
%figure traj_back_eps
% end figure traj_back_eps

\section{Astrophysical Applications}

\label{secapplication}
We now analyse two models for the synchrotron emission
along the jet for a Lane--Emden  ($n=5$) density profile.

\subsection{Direct Conversion}

The flux of observed radiation
along the centre of the jet, $I_c$,
in the classical case is  assumed to scale
as
\begin{equation}
I_c(x;x_0,v_0,b,\epsilon)\propto \frac{L(x;x_0,v_0,b,\epsilon)}{x^2}
\label{classicintensity}
\quad  ,
\end{equation}
where  $L$,
the sum of the radiative  losses
for a Lane--Emden  ($n=5$) density profile,
 is given by equation (\ref{classiclosses}).

This relation  connects  the observed
intensity of radiation with the rate of energy transfer per unit area.
In the relativistic case
\begin{equation}
I_c(x;x_0,\beta_0,b,c)\propto \frac{L_r(x;x_0,\beta_0,b,c)}{x^2}
\quad  ,
\label{relativisticintensity}
\end{equation}
where  $L_r$ is given by equation (\ref{relativisticlosses})

\begin{figure*}
\begin{center}
\includegraphics[width=4cm]{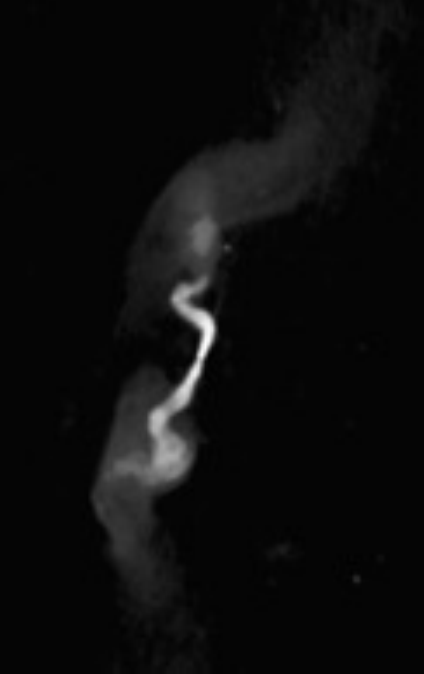}
\end {center}
\caption
{
Extended  image (300~kpc) of 3C31.
}
\label{gray_3c31}
    \end{figure*}

\begin{figure*}
\begin{center}
\includegraphics[width=8cm]{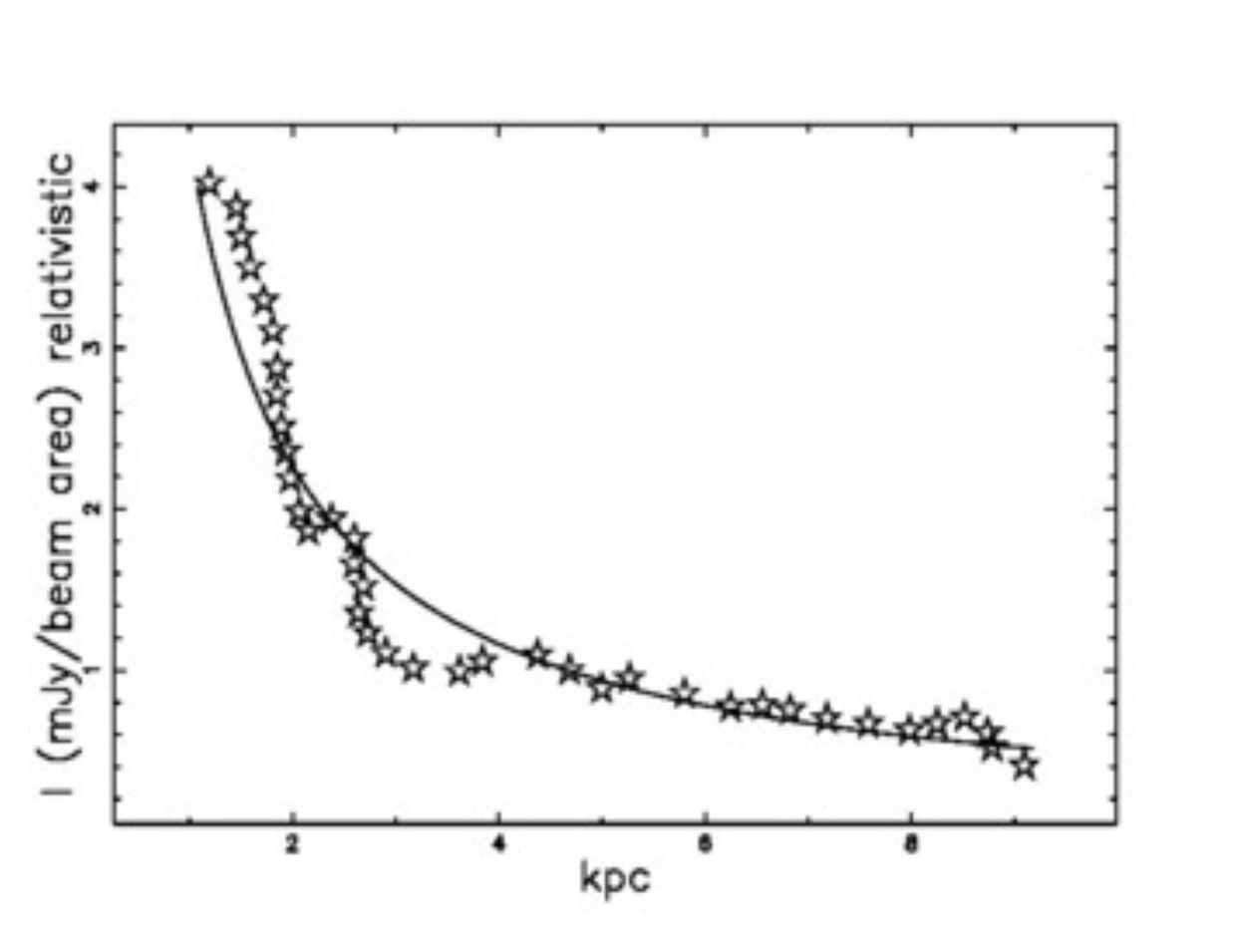}
\end {center}
\caption
{
Observed intensity profile along the centerline
of 3C31  (empty stars)
and  theoretical intensity
for a Lane--Emden  ($n=5$) density profile
as given
by equation (\ref{relativisticintensity}),
with  parameters as in Table \ref{jetrel} (full line).
The observational percentage  of
reliability is
$\epsilon_{\mathrm {obs}} =  86.19\%$.
}
\label{intensityrel_3c31}
    \end{figure*}

The
observational percentage  of
reliability can be used as a statistical test for the
the goodness of fit,
 $\epsilon_{\mathrm {obs}}$,
\begin{equation}
\epsilon_{\mathrm {obs}}  =100\bigl (1-\frac{ \sum_j |I_{obs}-I_{theo}|_j }
                                      { \sum_j I_{theo,j}           }
                              \bigr )
.
\label{efficiencymany}
\end{equation}
We now report the large scale structure and jets of 3C31,
see  Figure~\ref{gray_3c31}.
% figure   gray_3c31
% end gray_3c31
To make a comparison with the observed profile
of intensity, we choose  the first 10 kpc of 3C31,
see Figures 1 and  8 in
\cite{laing2002};
Figure \ref{intensityrel_3c31}
shows the theoretical synchrotron
intensity, as well as the observed one.

% figure   intensityrel_3c31
% end intensityrel_3c31

\subsection{The Magnetic Field of Equipartition}

\begin{figure*}
\begin{center}
\includegraphics[width=6.6cm]{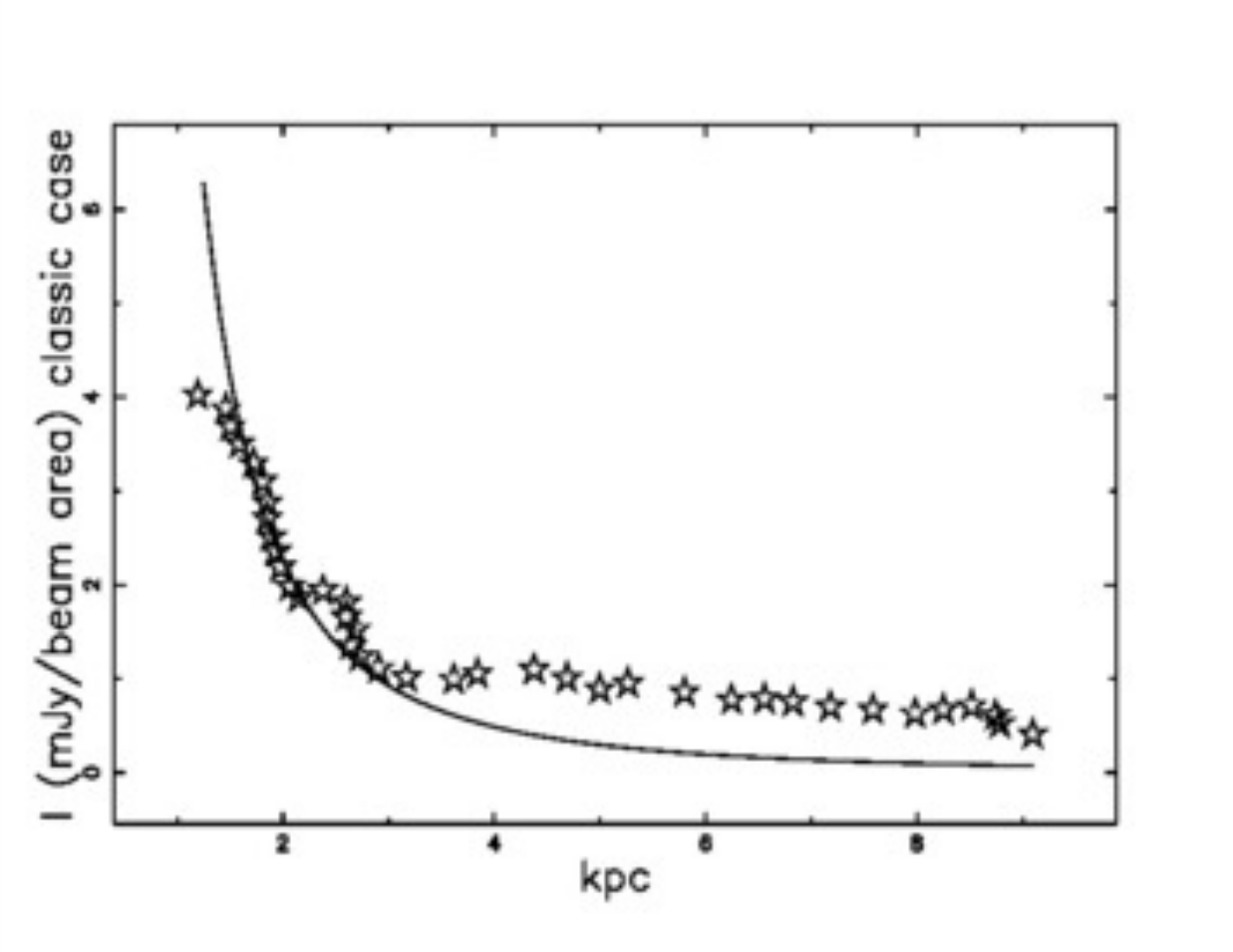}
\end {center}
\caption
{
Observed intensity profile along the centerline, $I_c$,
of 3C31  (empty stars)
and  theoretical intensity
for a Lane--Emden  ($n=5$) density profile
as given
by equation
(\ref{intensitycenterline}), with parameters
as in Table \ref{jetparameters}.
The observational percentage  of
reliability is
$\epsilon_{\mathrm {obs}} =  73.79\%$.
}
\label{intensitsynchro_3c31}
    \end{figure*}

The magnetic field  in CGS has an energy
density of $\frac{B^2}{8 \pi}$,
where $B$ is the magnetic field.
The presence of the magnetic field can be modeled
assuming equipartition between  the kinetic energy
and the magnetic energy
\begin{equation}
\frac{B(x)^2}{8 \pi} = \frac{1}{2} \rho v^2
\quad .
\end{equation}
By inserting the above equation
in the classical equation for the conservation
of the flux of energy
 (\ref{conservazioneenergy}), a factor 2
will appear on both sides of the equation, leaving
unchanged the result for the deduction of the velocity to the first order.
The magnetic field
as a function of the distance $x$
when  the velocity is  given by equation (\ref{vfirst})
and in the presence of a Lane--Emden ($n=5$) profile for the density
is
\begin{equation}
B(x;x_0,b) =
\frac
{
B_{{0}} \left( 3\,{b}^{2}+{x_{{0}}}^{2} \right) ^{{\frac{5}{12}}}{x_{{0
}}}^{{\frac{2}{3}}}
}
{
\left( 3\,{b}^{2}+{x}^{2} \right) ^{{\frac{5}{12}}}{x}^{{\frac{2}{3}}
}
}
\label{bx}
\quad .
\end{equation}
where $B_0$ is the magnetic field at $x=x_0$.
We  assume   an inverse power law  spectrum
for the ultrarelativistic  electrons,
of the type
\begin{equation}
N(E)dE = K E^{-p} dE
\label{spectrum}
\end{equation}
where $K$ is a constant and $p$ the exponent of the inverse power law.
The intensity of the synchrotron radiation has a standard
expression, as given
by formula (1.175)  in \cite{lang2},
\begin{eqnarray}
I(\nu)
\approx 0.933 \times 10^{-23}
\alpha_p (p) K l  H_{\perp} ^{(p +1)/2 }
\bigl (
 \frac{6.26 \times 10^{18} }{\nu}
\bigr )^{(p-1)/2 }  \\
erg\, sec^{-1} cm^{-2} Hz^{-1} rad^{-2}
\nonumber
\label{isynchro}
\end{eqnarray}
where $\nu$ is the frequency,
$H_{\perp}$ is the magnetic field perpendicular to the
electron's velocity,
$l$ is the dimension of the radiating region
along the line of sight,
and  $ \alpha_p (p)$  is a slowly
varying function
of $p$, which is of the order of unity.
We now analyse the intensity along the centerline of the jet,
which means that the  radiating length is
\begin{equation}
l(x;\alpha) = x   \tan ( \frac{\alpha}{2})
\quad .
\end{equation}
The intensity, assuming a constant $p$, scales as
\begin{equation}
I(x;x_0,p)=
\frac
{
I_{{0}}{B}^{{\frac {p}{2}}+{\frac{1}{2}}}x
}
{
{B_{{0}}}^{{\frac {p}{2}}+{\frac{1}{2}}}{\it x_0}
}
\quad ,
\end{equation}
where $I_0$  is the intensity at  $x=x_0$ and  $B_0$
the magnetic field
at $x=x_0$.
We insert Eq.~(\ref{bx})
to have an analytical
expression for the centerline intensity
\begin{equation}
I(x;x_0,p,b)=
 \left( 3\,{b}^{2}+{x_{{0}}}^{2} \right) ^{{\frac {5\,p}{24}}+{\frac{5
}{24}}}i_{{0}}{x}^{-{\frac {p}{3}}+{\frac{2}{3}}}{x_{{0}}}^{{\frac {p
}{3}}-{\frac{2}{3}}} \left( 3\,{b}^{2}+{x}^{2} \right) ^{-{\frac {5\,p
}{24}}-{\frac{5}{24}}}
\label{intensitycenter}
\quad .
\end{equation}
This equation for the intensity is relative to
the unit area;
to have the intensity on the centerline,
$I_c$,
we should make a further division by the area
of interest, which scales $\propto \,x^2$
\begin{equation}
I_c(x;x_0,p,b)= \frac{I(x;x_0,p,b)}{x^2}
\label{intensitycenterline}
\quad .
\end{equation}
Figure \ref{intensitsynchro_3c31}
shows the theoretical synchrotron
intensity with the variable magnetic field,
as well as the observed one for 3C31.

% figure   intensitsynchro_3c31

\section{Conclusion}

\subsection{ Classical Case}

We modeled the physics of turbulent jets by the
conservation of the energy flux.
In the case of constant density,
we derived solutions for the  distance and velocity
as functions of
time, see Eqs~(\ref{energyxtconstant}) and
(\ref{energyvtconstant}).
In the presence of  an hyperbolic profile of density,
the solutions for the  distance and velocity
as functions of
time are  Eqs~(\ref{xthyperbolic}) and
(\ref{vthyperbolic}).
The case of a density that follows an inverse power
law of density
is limited to the derivation of the velocity, see
Eq.~(\ref{velocitypower}).
The presence of an inverse power law introduces
flexibility in the results and, as an example,
when $\delta=2$
the rate of mass flow
does not increases  with $x$ but is constant; see
Eq.~(\ref{mxpower}).
The approximate trajectory
of a turbulent jet
in the presence of a  Lane--Emden ($n=5$) medium
has been evaluated to first order, see equation (\ref{xtasymptotic}).
The  solution for the velocity to first order
allows the  insertion of the back-reaction, i.e.,
the radiative losses, in the equation for the   flux of energy
conservation,
see equation (\ref{consfluxback}),
and consequently  the  velocity corrected to second order,
see equation (\ref{vcorrected}).
The  trajectory, calculated numerically  to the second order, is shown
in Figure  \ref{traj_back}.
The radiative losses allow us to evaluate the length  at which
the advancing velocity of the jet is zero.
This length  has a complicated analytical expression
and was presented numerically, see Figure \ref{xlunpc}.

\subsection {Relativistic Case}

The conservation of the relativistic energy flux  for turbulent
jets is analysed here in  three cases.
In the first case, we have a  surrounding medium
with  constant density and
the analytical result   is limited to
a series expansion for the solution, see Eq. (\ref{xtrelseries}).
In the second  case, the surrounding density
decreases with a power law behaviour and   the analytical result
is limited to the velocity--distance relation,
see Eq. (\ref{betadistance}),
and to a series expansion for the solution,
see Eq. (\ref{xtseriesreldelta}).
In the third   case, the surrounding density
decreases according to a Lane--Emden ($n=5$) medium
and  it is  possible to derive
an analytical expression for $\beta$ to the first order, see equation
(\ref{betarelfirst}), and
to the second order (taking into account radiative losses),
see equation
(\ref{betarelsecond}).
The relativistic  trajectory to the first order has been  evaluated
through  a series
for a Lane--Emden ($n=5$) medium, see equation (\ref{eqn_traj_rel_series})
or a Pad\'e  approximant of order [2/1],
see equation (\ref{eqn_traj_rel_pade}).
The relativistic equation of motion to second order (back-reaction)
has been  evaluated numerically,  see
Figure \ref{traj_back_eps}.
In other words, with the introduction of the radiative losses,
the length of the classical or relativistic
jet becomes finite rather than infinite.

\subsection {An Astrophysical Application}
The radiative losses  for a a Lane--Emden ($n=5$) medium are represented by
equation (\ref{lossesclassical})  in the classical case
and by (\ref{relativisticlosses}) in the relativistic case.
A division of these  two quantities by
the area
of interest allows us to derive
the theoretical rate of energy transfer per unit area,
which can be compared with the intensity
of radiation along the jet, for example, 3C31,
see Figure \ref{intensityrel_3c31}.
The  spatial behaviour of the magnetic field  is introduced under the hypothesis
of equipartition between the kinetic and magnetic energy,
see equation~(\ref{bx}), which  allows us to close the
standard equation for the synchrotron emissivity, see equation
(\ref{isynchro}).

%\bibliography{biblio}

\label{lastpage-01}
\printindex
\end{document}